\documentclass[reprint,superscriptaddress,amsmath,amssymb,aps,showpacs]{revtex4-1}
\usepackage[utf8]{inputenc}
\usepackage{graphicx}
\usepackage{bm}
\usepackage{epsfig}
\usepackage{amssymb}
\usepackage{amsfonts}
\usepackage{color}
\usepackage{natbib}
\usepackage[caption=false]{subfig}

\graphicspath{{figures/}}
\usepackage{caption}

\begin{document}

\title{Half metals at intermediate energy scales in Anderson-type insulators}

\date{\today}

\author{Kyung-Yong Park}
\affiliation{Department of Physics, POSTECH, Pohang, Gyeongbuk 37673, Korea}

\author{Hyun-Jung Lee}
\affiliation{Seoul Science High School, Seoul 03066, Korea}

\author{Ki-Seok Kim}
\affiliation{Department of Physics, POSTECH, Pohang, Gyeongbuk 37673, Korea}
\affiliation{Asia Pacific Center for Theoretical Physics (APCTP), Pohang, Gyeongbuk 37673, Korea}

\begin{abstract}
Although quantum phase transitions involved with Anderson localization had been investigated for more than a half century, the role of spin polarization in these metal-insulator transitions has not been clearly addressed as a function of both the range of interactions and energy scales. Based on the Anderson-Hartree-Fock study, we reveal that the spin polarization has nothing to do with Anderson metal-insulator transitions in three dimensions as far as effective interactions between electrons are long-ranged Coulomb type. On the other hand, we find that metal-insulator transitions appear with magnetism in the case of Hubbard-type local interactions. In particular, we show that the multifractal spectrum of spin $\uparrow$ electrons differs from that of spin $\downarrow$ at the high-energy mobility edge, which indicates the existence of spin-dependent universality classes for metal-insulator transitions. One of the most fascinating and rather unexpected results is the appearance of half metals at intermediate energy scales above the high-energy mobility edge in Anderson-type insulators of the Fermi energy, that is, only spin $\uparrow$ electrons are delocalized while spin $\downarrow$ electrons are Anderson-type localized.
\end{abstract}

\maketitle

\section{Introduction}

Since metal-insulator transitions had been observed in two dimensions, there have appeared both extensive and intensive studies on the interplay between Anderson localization and electron correlation
%
%
\cite{2DMIT_Kravchenko_Review}. On the other hand, the electronic structure near Anderson metal-insulator transitions, particularly involved with spin degrees of freedom, has not been characterized in three dimensions, compared to the case of two-dimensional metal-insulator transitions. Recent experiments revealed that eigenfunction multifractality \cite{AMIT_Review} still survives in the presence of Coulomb interactions, where the measured multifractal exponent is enhanced and the nature of multifractality becomes weakened \cite{Multifractality_Interaction_Experiment}. These experiments promoted theoretical investigations such as nonlinear $\sigma-$model field-theory study for multifractal exponents \cite{Multifractality_Interaction_NLsM} and metal-insulator transitions based on the Hartree-Fock approximation for interactions \cite{Multifractality_Interaction_Numerics} and in hybridization with density functional theory \cite{Coulomb_Disorder_DFT_I,Coulomb_Disorder_DFT_II}. Recently, two of the authors showed that two types of mobility edges exist in the presence of long-ranged Coulomb interactions \cite{Two_Mobility_Edges}. The multifractal analysis within the Hartree-Fock approximation confirms that the high-energy mobility edge is connected to that of the Anderson transition without interactions while the low-energy mobility edge near the Fermi energy occurs from the interplay between Coulomb correlation and Anderson localization. In particular, it turns out that each mobility edge is described by two different multifractal spectral functions, which indicates that the respective metal-insulator transitions near the Fermi energy and at the high-energy mobility edge belong to two different universality classes for quantum phase transitions \cite{Two_Mobility_Edges}.

Although the interplay between Anderson localization and electron correlation has been shown to result in a novel electronic structure for three-dimensional metal-insulator transitions, there still remains an essential question, that is, the role of spin degrees of freedom in these metal-insulator transitions. Resorting to the multifractal analysis within the Anderson-Hartree-Fock study, we examine how local magnetization affects these metal-insulator transitions as a function of both the range of electron correlations and energy scales. First, we reveal that the local spin polarization has nothing to do with Anderson metal-insulator transitions in three dimensions when effective interactions between electrons are long-ranged Coulomb type. More precisely, both magnetism and their local fluctuations appear deep inside an insulating phase. Second, on the other hand, we find that metal-insulator transitions occur with magnetism and their local fluctuations in the case of Hubbard-type local interactions. In particular, we show that the multifractal spectrum of spin $\uparrow$ electrons differs from that of spin $\downarrow$ at the high-energy mobility edge, which indicates the existence of spin-dependent universality classes for metal-insulator transitions. One of the most fascinating and rather unexpected results is the appearance of half metals at intermediate energy scales above the high-energy mobility edge in Anderson-type insulators of the Fermi energy, that is, only spin $\uparrow$ electrons are delocalized while spin $\downarrow$ electrons are Anderson-type localized.

\section{Model Hamiltonian and Hartree-Fock-Anderson analysis}

We consider an effective model Hamiltonian on a three-dimensional cubic lattice, where the quadratic part is
\begin{align}
    H_0 = - t \sum_{\langle i j \rangle} \sum_{\sigma} c_{i \sigma}^{\dagger} c_{j \sigma} + h. c. + \sum_{i} \sum_{\sigma} (\epsilon_i - \mu) c_{i \sigma}^{\dagger} c_{i \sigma} ,
\end{align}
and the interaction term is
\begin{align}
    H_I = \frac{1}{2} \sum_{i} \sum_{j} \sum_{\sigma \sigma'} c_{i \sigma}^{\dagger}c_{i \sigma} U_{ij}^{\sigma\sigma'} c_{j \sigma'}^{\dagger}c_{j \sigma'}.
\end{align}
Here, $t$ is the hopping integral between nearest neighbor sites $\langle i j \rangle$, and $\mu$ is the chemical potential. In this study we set $t = 1$ as the unit of energy and focus on the case of half filling. $\epsilon_i$ is a random potential, uniformly distributed in $[- W, W]$. It is well known that the Anderson transition occurs at $W_{c} = 8.25$ without electron correlations \cite{AMIT_Critical_Disorder_Strength}. Here, we set $W = 7 < W_c$. The interaction coefficient is given by $U_{ij}^{\sigma\sigma'} = \frac{U_{\sigma \sigma'}}{r_{ij}}$ with the strength $U_{\sigma \sigma'} \equiv \frac{e^2}{\kappa} \sigma \sigma'$ for the case of Coulomb and $U_{ij}^{\sigma\sigma'} = U \delta_{- \sigma \sigma'} \delta_{ij}$ for the case of Hubbard, respectively. $\sigma$ represents $+$ ($-$) for spin $\uparrow$ ($\downarrow$) electrons.

Resorting to the Hartree-Fock approximation for interactions, we perform the self-consistent analysis in real space for both (i) spin-density and charge-density order parameters of the Hartree channel and (ii) exchange hopping order parameters of the Fock channel, and find corresponding eigenfunctions and eigen energies for a given disorder configuration. Then, we take averaging for various disorder configurations in a brute force way: Twenty samples for disorder realizations in the long-ranged interacting system (LRIS) and hundred ones in the short-ranged interacting system (SRIS). We would like to emphasize that the Ewald-summation technique \cite{Ewald_Sum_Technique1,Ewald_Sum_Technique2,Ewald_Sum_Technique3} has been utilized, which allows us to deal with long-ranged Coulomb interactions quite accurately, where the long-ranged part of interactions is taken in the momentum space while the short-ranged interaction is considered in the real space \cite{Comment_Ewald_Sum}. See Ref. \cite{Two_Mobility_Edges} for more details. In addition, we perform our simulations in the three dimensional cubic lattice, varying the system size $L^{3}$ with $L = 12$, $16$, and $20$ for the multifractal analysis to identify the mobility edge.

It is not easy to justify the Hartree-Fock approximation, in particular, for the regime of strong correlations, where Mott physics is involved. In appendix A, (i) we argue that the Hartree-Fock-Anderson method (real-space Hartree-Fock computation with brute force disorder averaging) should work well in the weak coupling regime beyond the Finkelstein's renormalization group analysis \cite{Finkelstein_RG} (regarded to be the standard theoretical framework in the weak coupling regime) not only in the metallic regime but also in the insulating phase \cite{Justification_for_i}, and (ii) comparing our Hartree-Fock-Anderson analysis with Ref. \cite{Henseler}, the study of which tried to cover the Mott regime of strong correlations, we confirm that the half metallic state appears robustly in the weak coupling regime.

\section{Appearance of half metals at intermediate energy scales in Anderson-type insulators}

\subsection{Density of states and magnetization}

\begin{figure}
\includegraphics[scale=0.25]{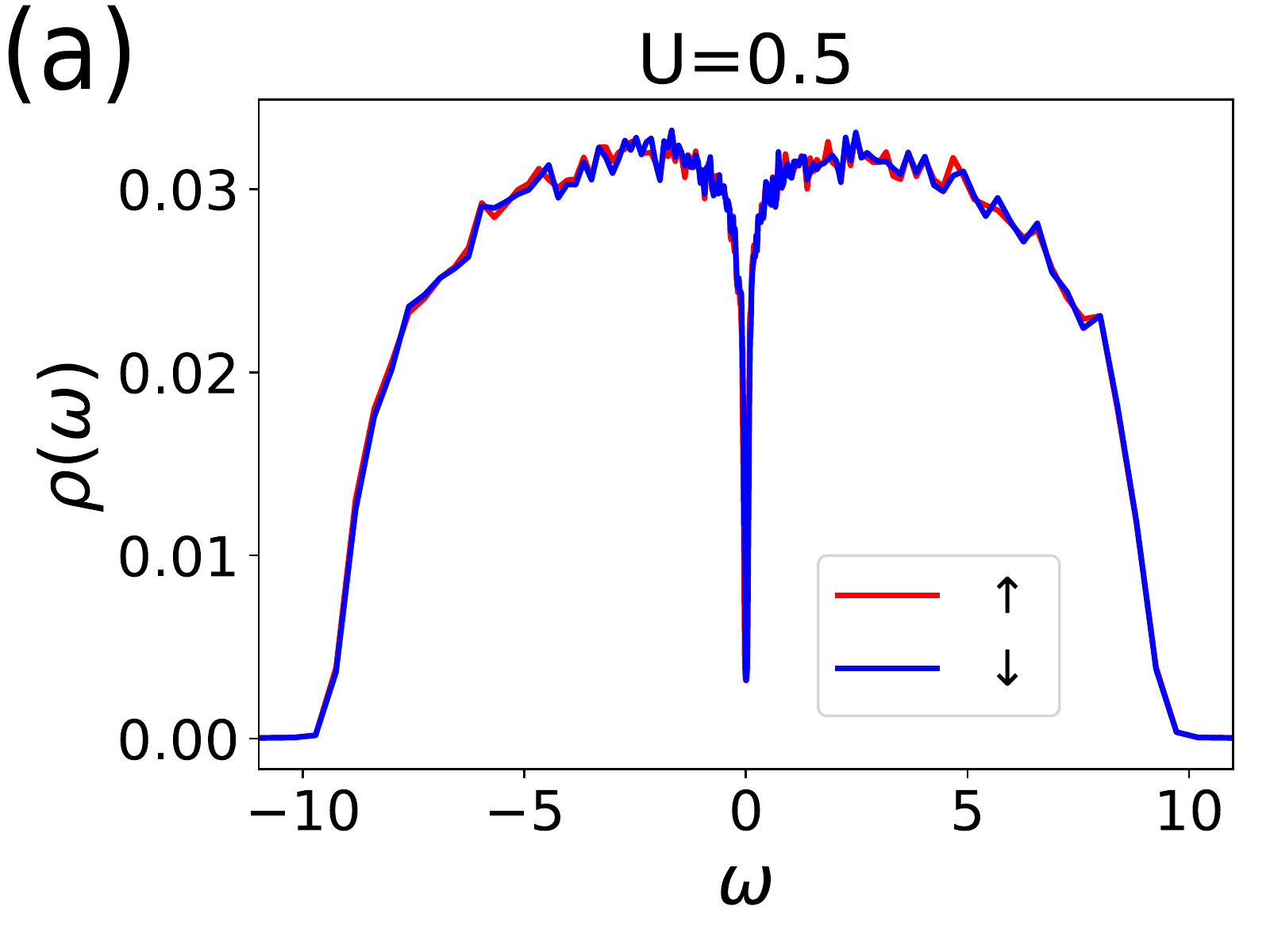}
\includegraphics[scale=0.25]{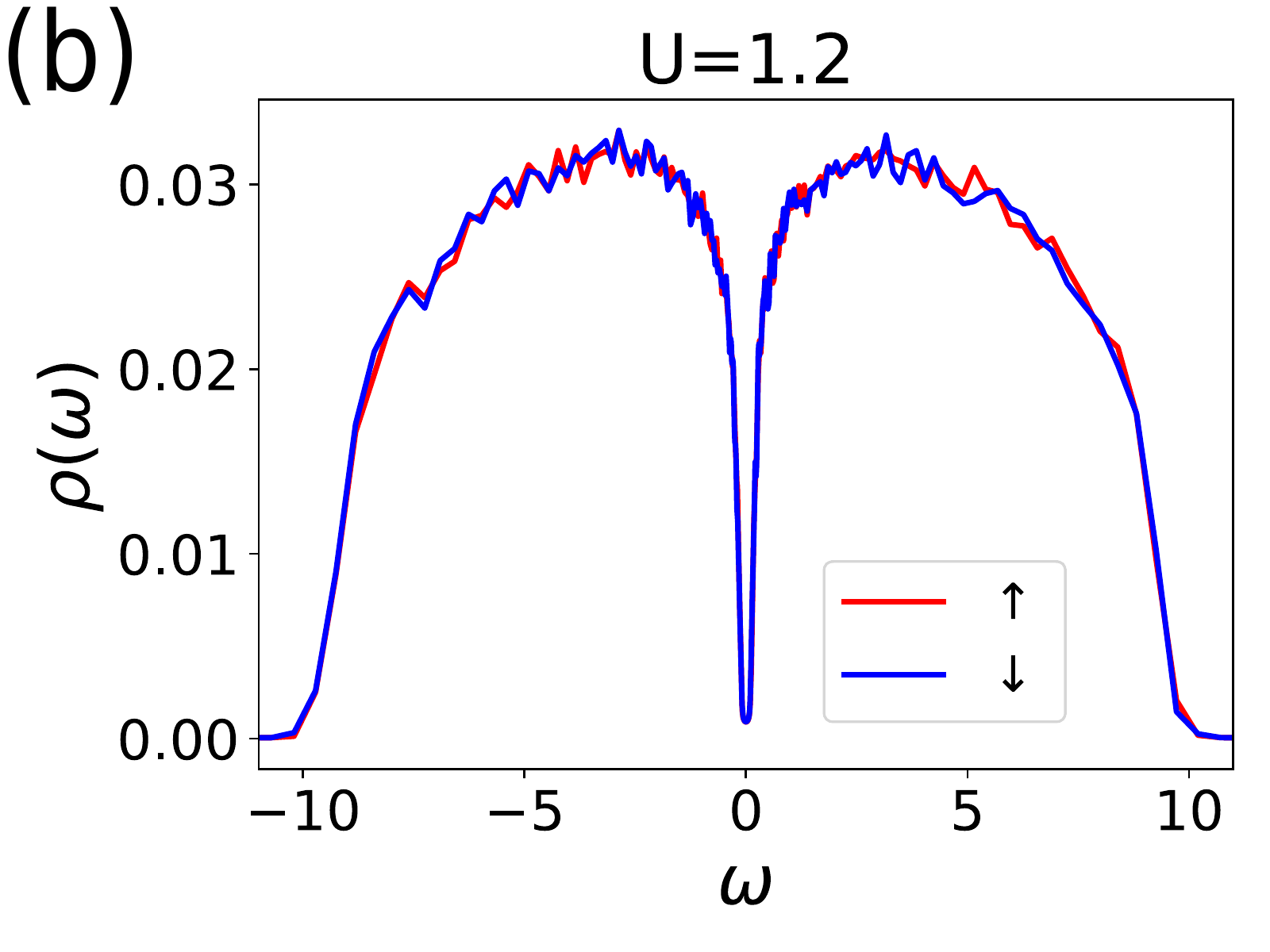}
  \caption{Disorder averaged density of states (DOS) $\rho(\omega)$ for various interaction parameters in the long-ranged interacting system (LRIS).
  This spin-resolved DOS does not show much difference in this interaction range, compared with that of the spinless case \cite{Two_Mobility_Edges}.
  The dip feature of the DOS at the Fermi energy is already pronounced in this interaction range, identified with the Coulomb gap in the log-log plot and not shown here but carefully discussed in the spinless case. According to the multifractal analysis in the spinless case, the metal-insulator transition at the Fermi energy occurs around $U \approx 0.3$ in the absence of any magnetic ordering, confirmed by the present multifractal analysis.}
 \label{fig:DOS_LRIS}
\end{figure}

\begin{figure}
\includegraphics[scale=0.25]{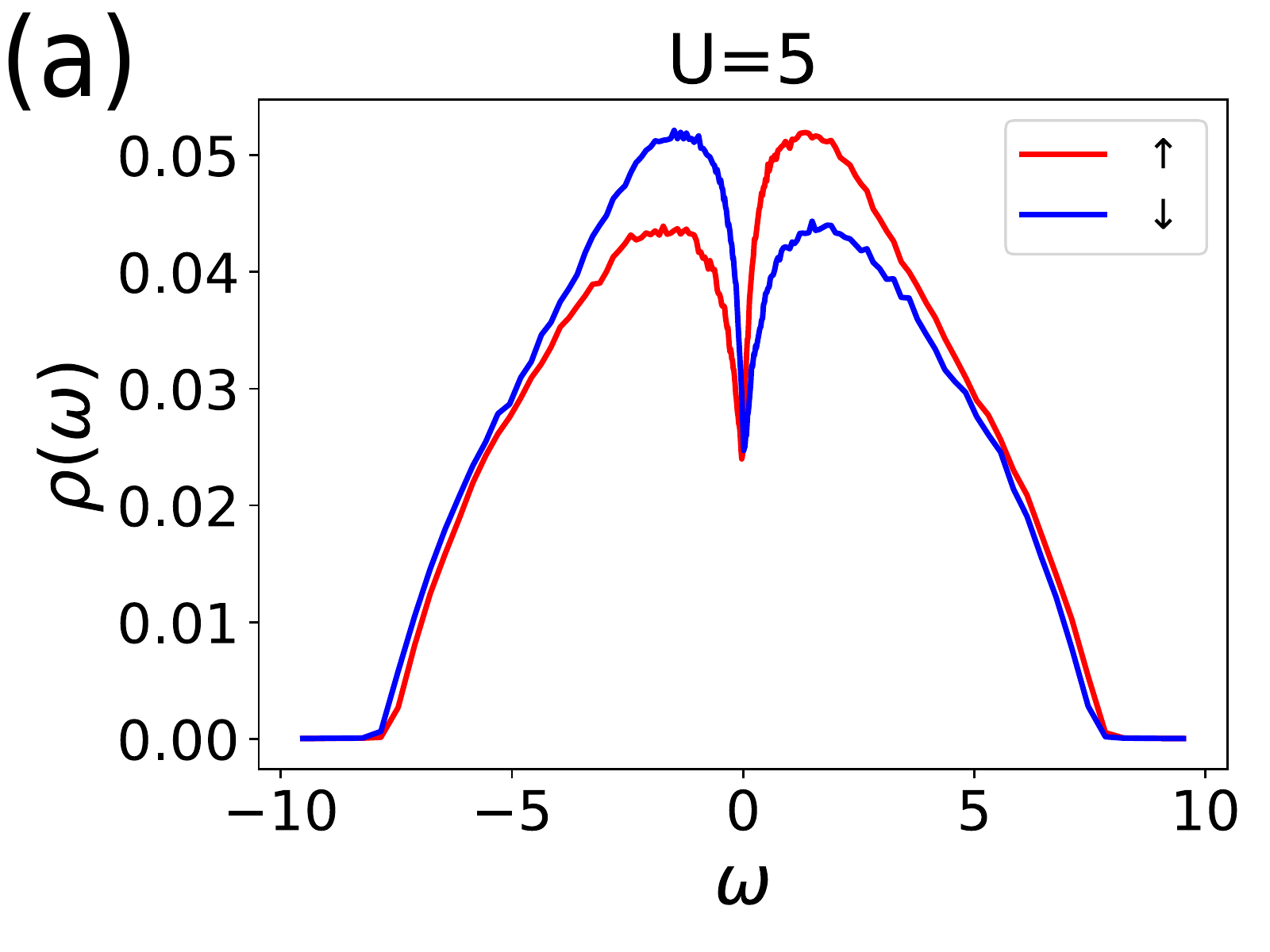}
\includegraphics[scale=0.25]{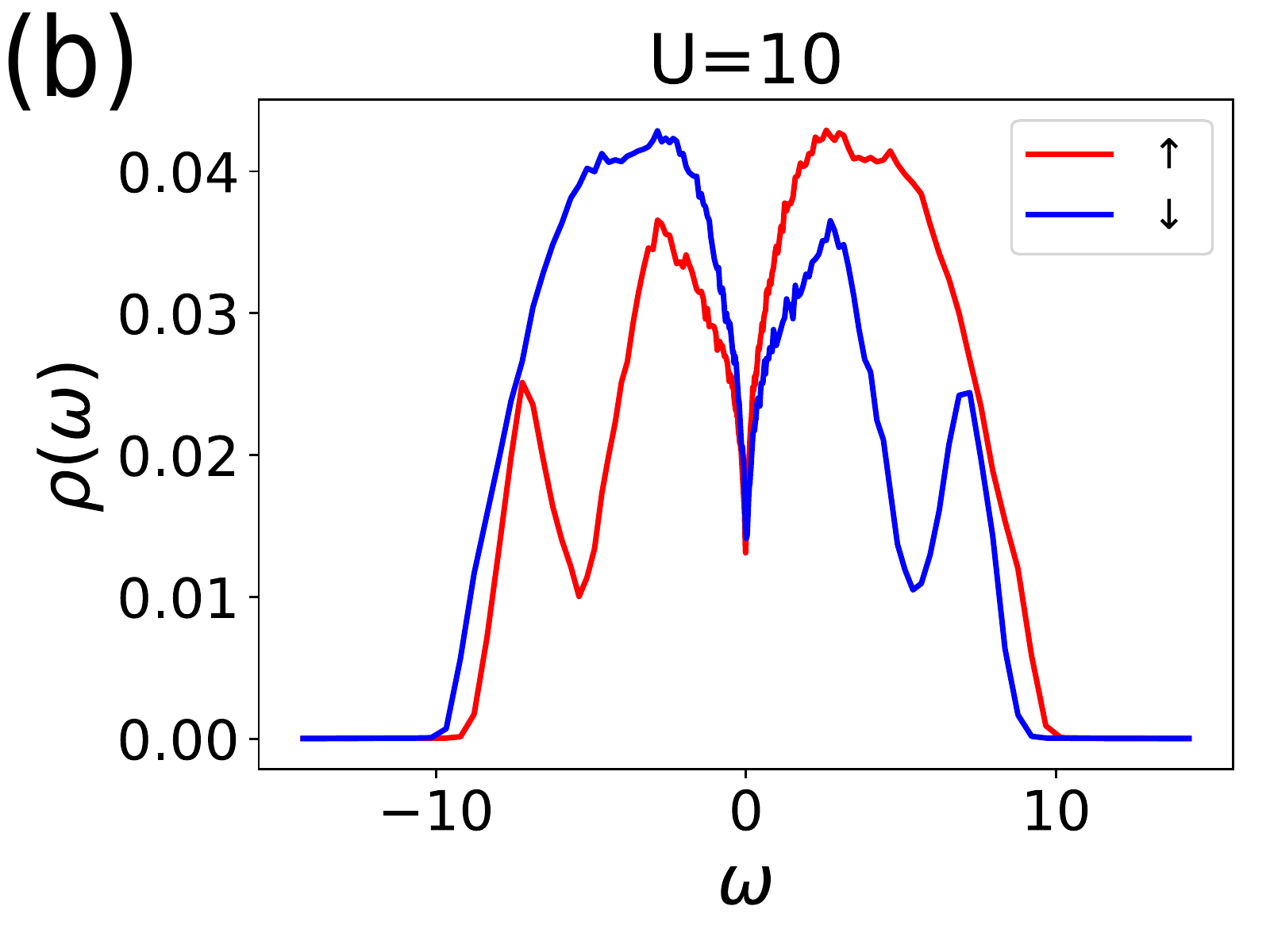}

  \caption{Disorder averaged density of states (DOS) $\rho(\omega)$ for various interaction parameters in the short-ranged interacting system (SRIS).
  The DOS evolution of the SRIS shows strong spin resolution in contrast with that of the LRIS. Besides the dip feature at the Fermi energy, there appears additional suppression of the DOS at an intermediate energy scale, identified with $\omega_{U}$, here. This suppressed DOS is transferred to enhance the DOS at the other energy scale of $\omega_{U} + U$. The opposite-spin DOS shows the opposite behavior, that is, enhancement of the DOS at $\omega_{U}$ but its suppression at $\omega_{U} + U$.}
 \label{fig:DOS_SRIS}
\end{figure}

Disorder averaged density of states (DOS) $\rho(\omega)$ are shown in Fig. \ref{fig:DOS_LRIS} (LRIS) and Fig. \ref{fig:DOS_SRIS} (SRIS). First of all, it turns out that the spin-resolved DOS remains essentially the same as that of the spinless case \cite{Two_Mobility_Edges} in the LRIS. The dip feature of the DOS at the Fermi energy is already pronounced in the interaction range of $0.3 \leq U \leq 1.2$,
%
%
identified with the Efros-Shklovskii pseudogap \cite{Coulomb_Gap_Review} in the log-log plot \cite{McMillan_Shklovskii_scaling_theory} (not shown here but carefully discussed in the spinless case \cite{Two_Mobility_Edges}). We recall $U = \frac{e^2}{\kappa}$ in the Coulomb case. According to the multifractal analysis in the spinless case, the metal-insulator transition at the Fermi energy occurs around $U \approx 0.3$, confirmed by the present multifractal analysis below. On the other hand, the DOS evolution of the SRIS shows strong spin resolution in contrast with that of the LRIS. Besides the dip feature at the Fermi energy, where the Altshuler-Aronov correction behavior \cite{Altshuer_Aronov_Correction} is mixed with the Efros-Shklovskii pseudogap feature \cite{Coulomb_Gap_Review} in the log-log plot and not shown here, there appears additional suppression of the DOS at an intermediate energy scale, identified with $\omega_{U}$. This suppressed DOS is transferred to enhance the DOS at the other energy scale of $\omega_{U} + U$. The opposite-spin DOS shows the opposite behavior, that is, enhancement of the DOS at $\omega_{U}$ but its suppression at $\omega_{U} + U$. This spin-resolved DOS suppression and enhancement at each intermediate energy scale is one of the main discoveries in this study.

%
%

\begin{figure}
\includegraphics[scale=0.25]{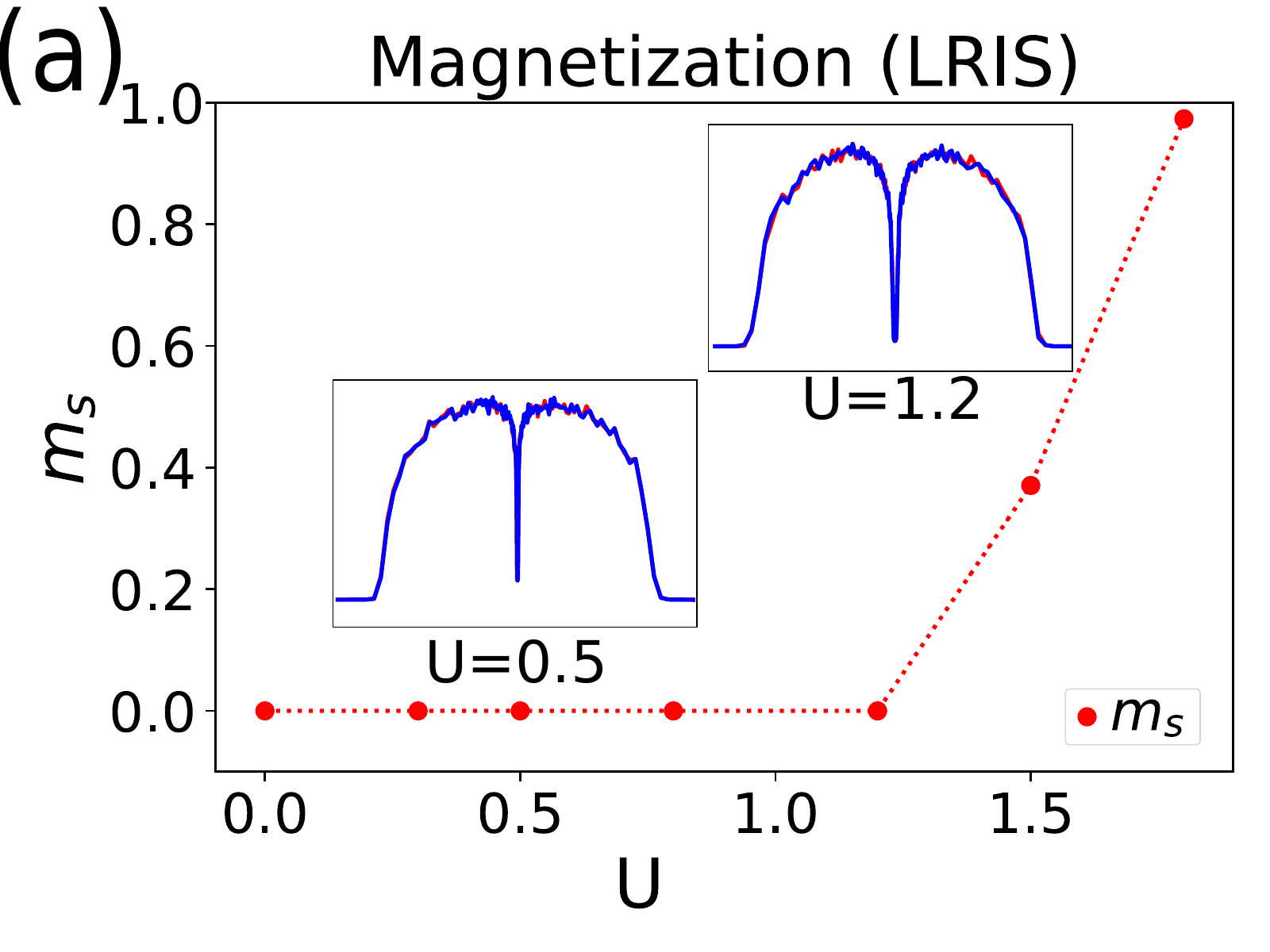}
\includegraphics[scale=0.25]{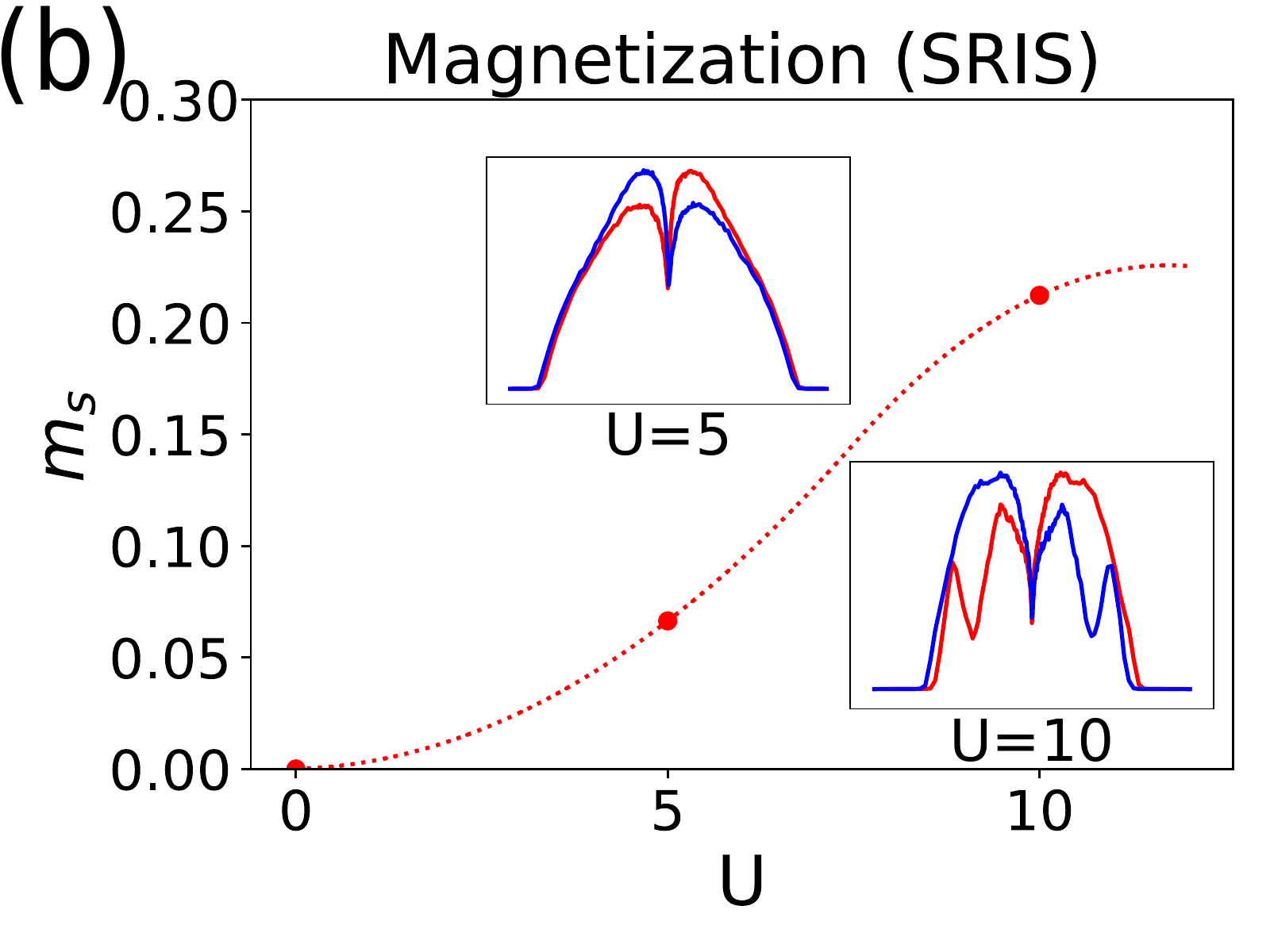}
  \caption{Disorder averaged magnetization as a function of the interaction strength in both LRIS and SRIS. Magnetization appears after the metal-insulator transition in the LRIS while ferromagnetism already exists before the metal-insulator transition in the SRIS. Magnetization appears at $1.2 < U < 1.5$ in the LRIS while the metal-insulator transition occurs at around $0.3$. We also check out fluctuations of local magnetization, varying the disorder realization, and find that such local fluctuations of magnetism are weak enough to be neglected in the regime of the metal-insulator transition, not shown here. On the other hand, ferromagnetism already occurs near the metal-insulator transition in the SRIS, the critical interaction strength of which is $5 < U < 10$, given by the multifractal analysis. In addition, we observed strong spatial fluctuations of the local magnetization in the metal-insulator transition, not shown here. Inset figures show the spin-resolved local DOS of Figs. \ref{fig:DOS_LRIS} and \ref{fig:DOS_SRIS}.}
  \label{fig:Magnetization_LRIS_SRIS}
\end{figure}

Figure \ref{fig:Magnetization_LRIS_SRIS} shows disorder averaged magnetization as a function of the interaction strength in both LRIS and SRIS. Magnetization appears at $1.2 < U < 1.5$ in the LRIS while the metal-insulator transition occurs at around $0.3$ \cite{Two_Mobility_Edges}. We also check out fluctuations of local magnetization, varying the disorder realization, and find that such local fluctuations of magnetism are weak enough to be neglected in the regime of the metal-insulator transition. On the other hand, ferromagnetism already occurs near the metal-insulator transition in the SRIS, the critical interaction strength of which is $5 < U < 10$, given by the below multifractal analysis. In addition, we observed strong spatial fluctuations of the local magnetization in the metal-insulator transition. This comparison study leads us to conclude that the spin dynamics has nothing to do with the metal-insulator transition in the LRIS while local magnetic fluctuations would affect the Anderson transition in the SRIS.

\subsection{Multifractal Analysis and multifractal spectrum}

Strong fluctuations of eigenfunctions are one of the fingerprints near the Anderson metal-insulator transition, responsible for the nature of eigenfunction multifractlity \cite{AMIT_Review}. This eigenfunction multifractality can be characterized by the fractal dimension of each moment of an eigenfunction, given by a set of inverse participation ratios and their disorder averages. To identify the mobility edge as a function of the interaction strength, one obtains the multifractal scaling exponent from the fractal dimension by differentiating it with respect to the number of the eigenfunction moment \cite{Two_Mobility_Edges}. We follow the standard procedure for actual numerical calculations \cite{Chhabra1989,Janssen1994}, well known in this research and not shown here. An essential feature of the multifractal scaling exponent ($\alpha_q$) is that it is defined at each eigenenergy and becomes enhanced in metals as the size of the system is enlarged. Here, we focus on the case of positive $q$. On the other hand, it decreases in insulators as the volume is expanded. Based on this scaling behavior, one may expect that the multifractal scaling exponent would be unchanged regardless of the system size, which determines the mobility edge. However, it turns out that the effect of irrelevant corrections to scaling plays a key role in the description of the Anderson transition \cite{MFM_Numerics}. Actually, the position of the mobility edge bears a system-size dependent component. To determine the mobility edge precisely, Ref. \cite{MFM_Numerics} proposed a way of ``multifractal finite-size scaling", which introduces irrelevant scaling variables into the finite-size scaling analysis for the multifractal scaling exponent. We point out that this is quite a delicate procedure, not performed in the present study. In this respect our multifractal scaling exponents and their corresponding multifractal spectrum are rather qualitative than quantitative. Instead, the main point is the spin-polarization dependence of the multifractal scaling exponent, which implies spin dependent universality classes for metal-insulator transitions.

\begin{figure}
\includegraphics[scale=0.25]{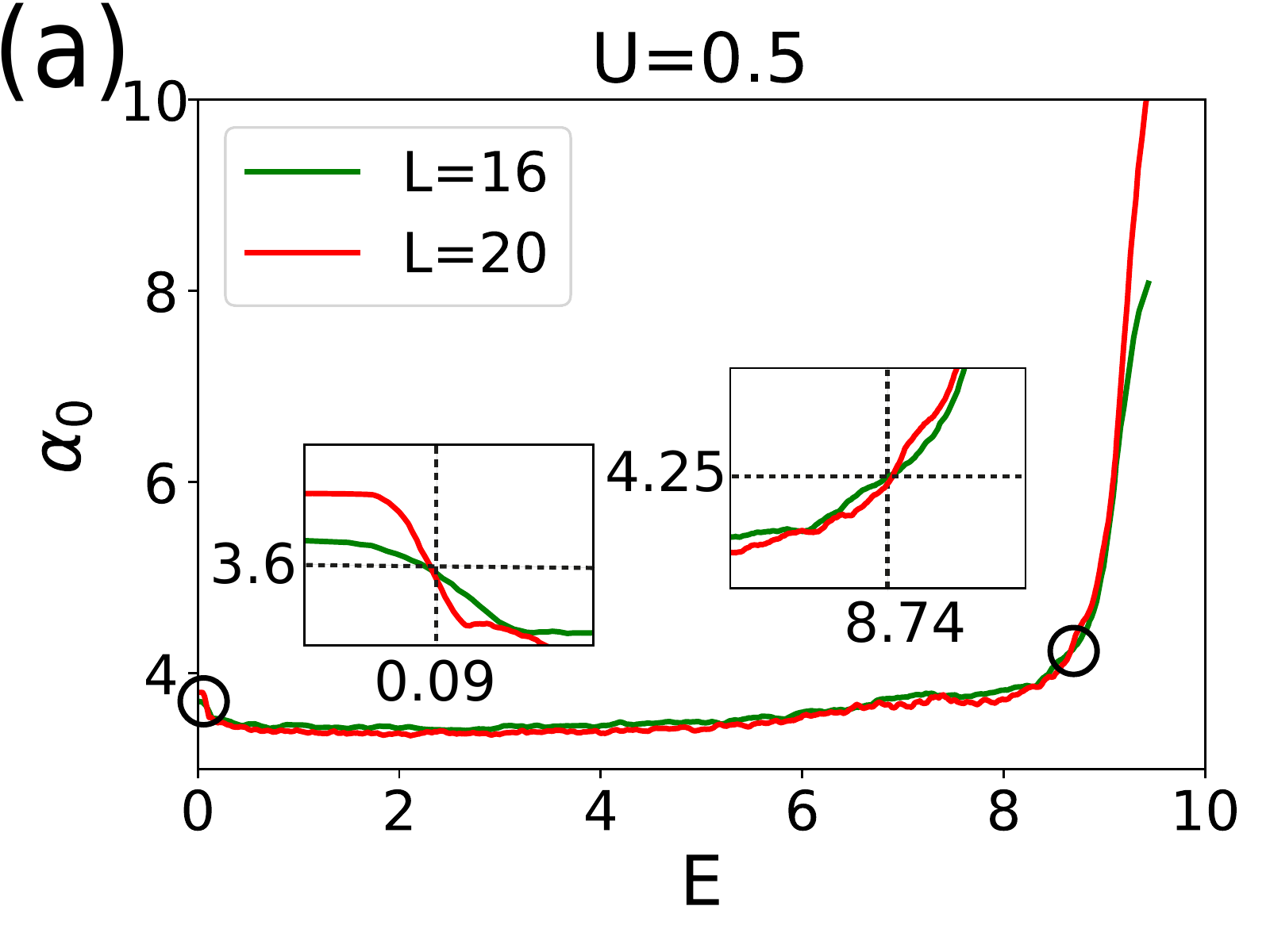}
\includegraphics[scale=0.25]{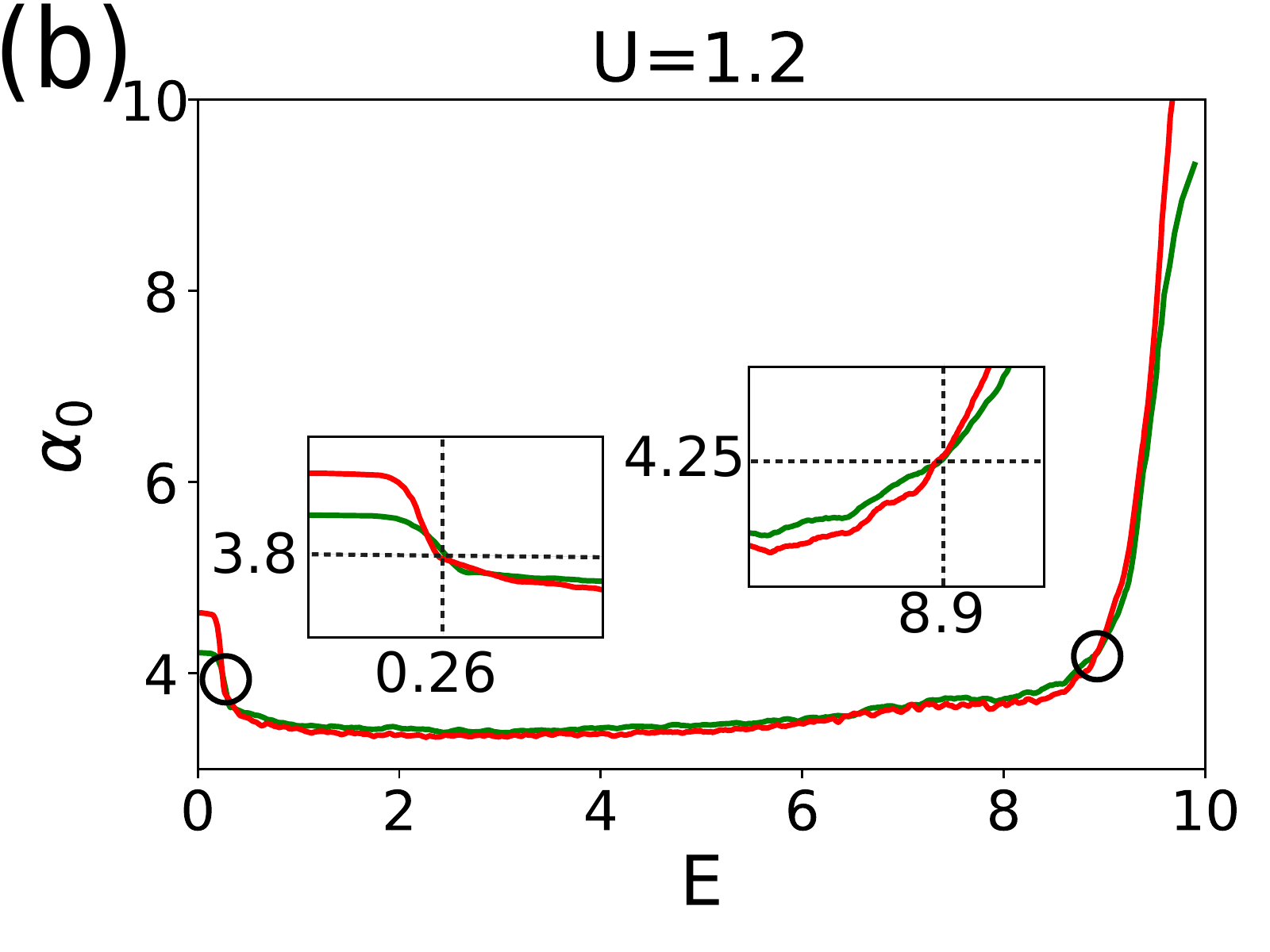}
\includegraphics[scale=0.25]{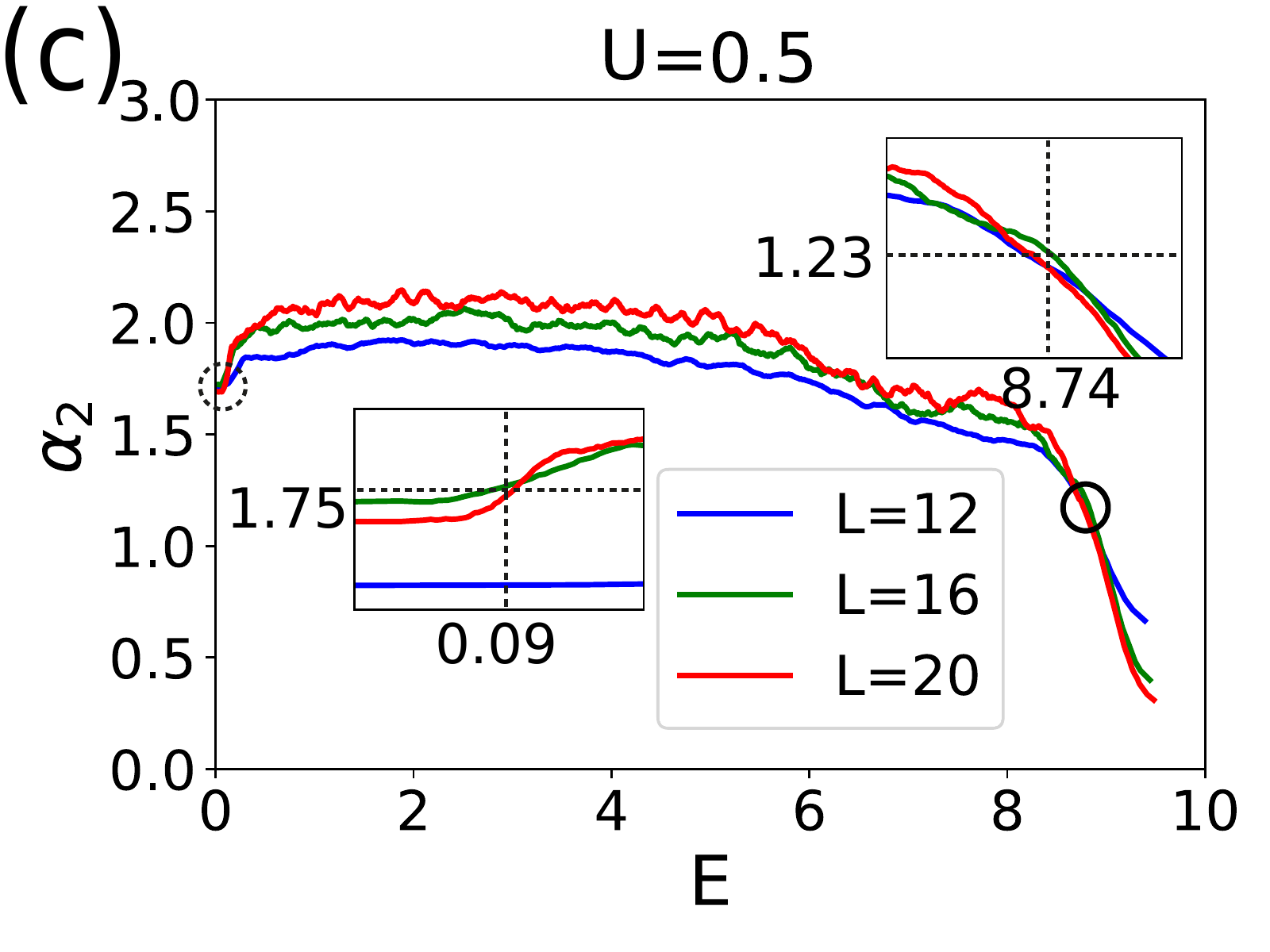}
\includegraphics[scale=0.25]{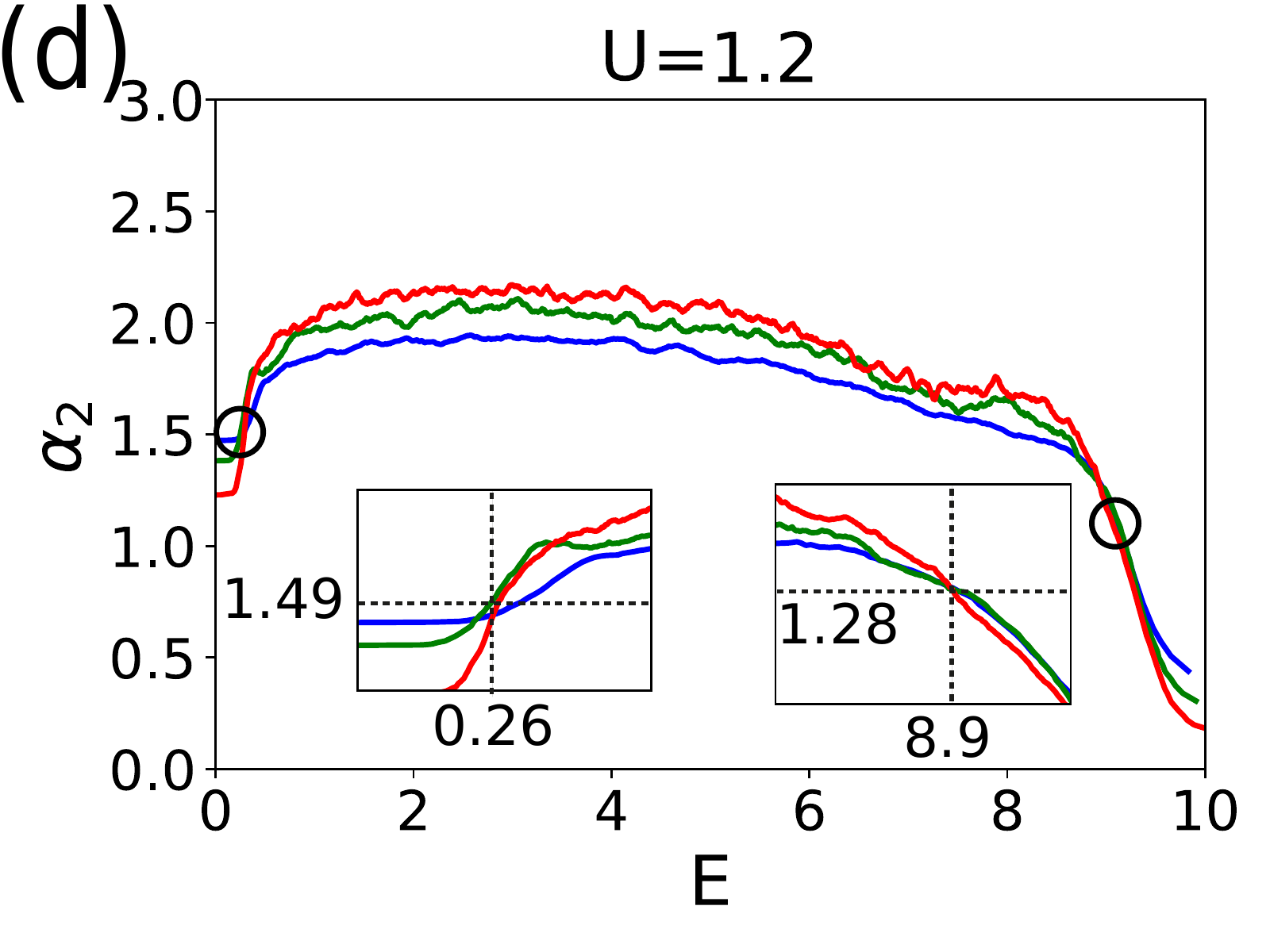}
  \caption{Spin-resolved multifractal scaling exponents $\alpha_0$ for $L = 16, ~ 20$ and $\alpha_{2}$ for $L = 12, ~ 16, ~ 20$ in the LRIS. As shown in the DOS of Fig. \ref{fig:DOS_LRIS}, the multifractal scaling exponent remains unchanged from that of the spinless case \cite{Two_Mobility_Edges}, where $\alpha_0^{\uparrow}$ and $\alpha_0^{\downarrow}$ ($\alpha_2^{\uparrow}$ and $\alpha_2^{\downarrow}$) are almost identical. Here, we marked two types of mobility edges, more clearly shown in the insets, where the high-energy mobility edge is connected to that of the Anderson transition without interactions while the low-energy mobility edge near the Fermi energy occurs from the interplay between Coulomb correlation and Anderson localization.}
  \label{fig:Multifractal_Scaling_Exponent_LRIS}
\end{figure}

We show spin-resolved multifractal scaling exponents $\alpha_0$ for two system sizes of $L =16$ and $20$ and $\alpha_{2}$ for $L = 12, ~ 16, ~ 20$ in the LRIS, given by Fig. \ref{fig:Multifractal_Scaling_Exponent_LRIS}. As shown in the DOS of Fig. \ref{fig:DOS_LRIS}, the multifractal scaling exponent remains unchanged from that of the spinless case, where $\alpha_0^{\uparrow}$ and $\alpha_0^{\downarrow}$ ($\alpha_2^{\uparrow}$ and $\alpha_2^{\downarrow}$) are almost identical. Here, we marked `size-independent' two crossing points for scale-invariant multifractal exponents \cite{MFM_Numerics}, more clearly shown in the insets, which identify two types of mobility edges. It turns out that the high-energy mobility edge is connected to that of the Anderson transition without interactions while the low-energy mobility edge near the Fermi energy occurs from the interplay between Coulomb correlation and Anderson localization \cite{Two_Mobility_Edges}. The low-energy crossing point close to the Fermi energy starts to appear from around $U = 0.3$, and the multifractal scaling exponent $\alpha_0$ saturates to the value of the spinless case around $U = 0.8$, from which it does not depend on the interaction strength, consistent with the previous study \cite{Two_Mobility_Edges}. This leads us to conclude that the critical strength for the Anderson metal-insulator transition occurs at $0.3 \leq U_c^{\mathrm{MIT}} < 0.5$. We recall the critical strength $1.2 < U_c^{\mathrm{FM}} < 1.5$ for ferromagnetism. Even though our Anderson-Hartree-Fock analysis overestimates the strength of ferromagnetism, $U_c^{MIT} \ll U_c^{FM}$ confirms that the local ferromagnetism has nothing to do with the Anderson metal-insulator transition in the LRIS.

\begin{figure}
\includegraphics[scale=0.25]{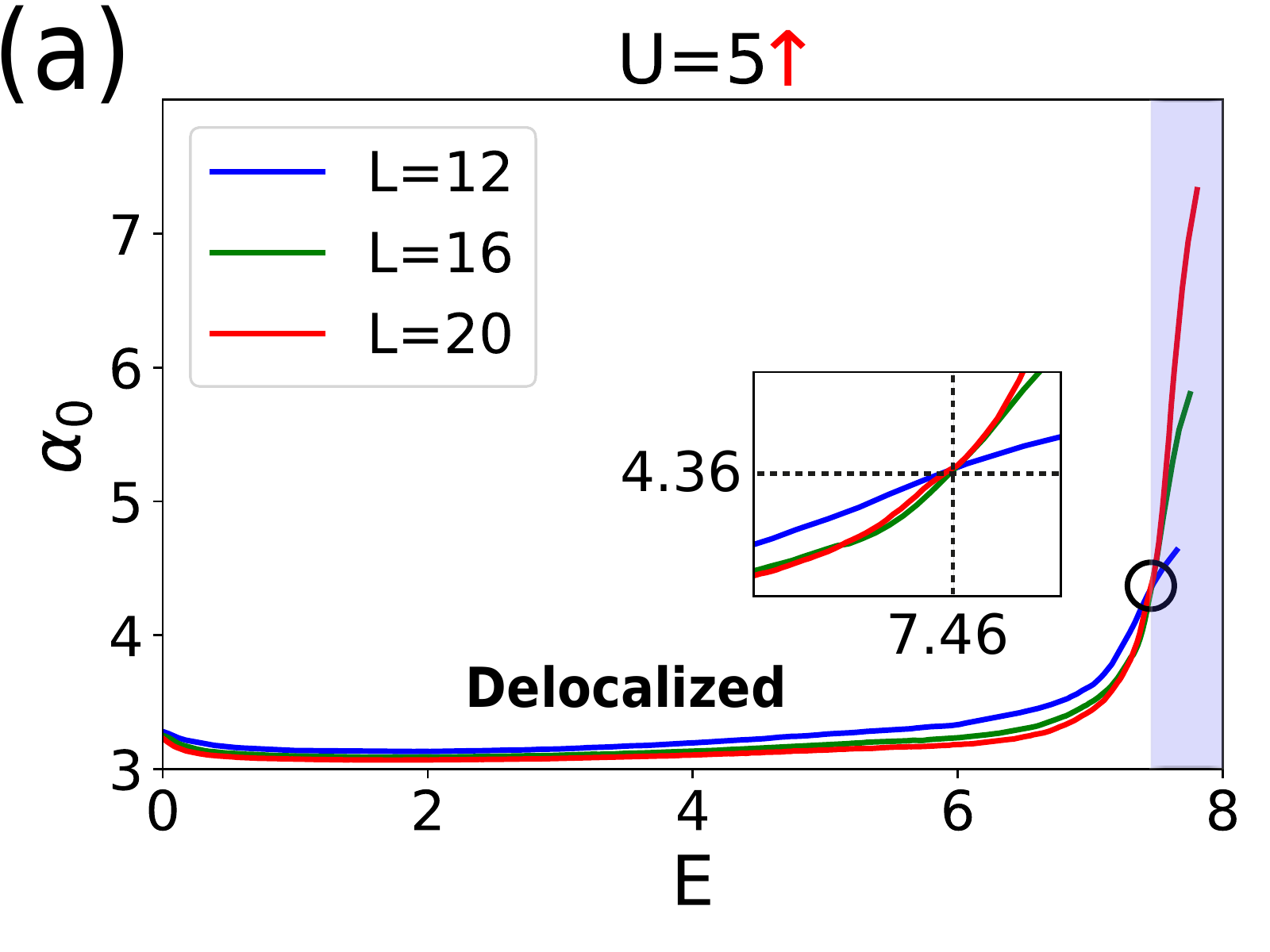}
\includegraphics[scale=0.25]{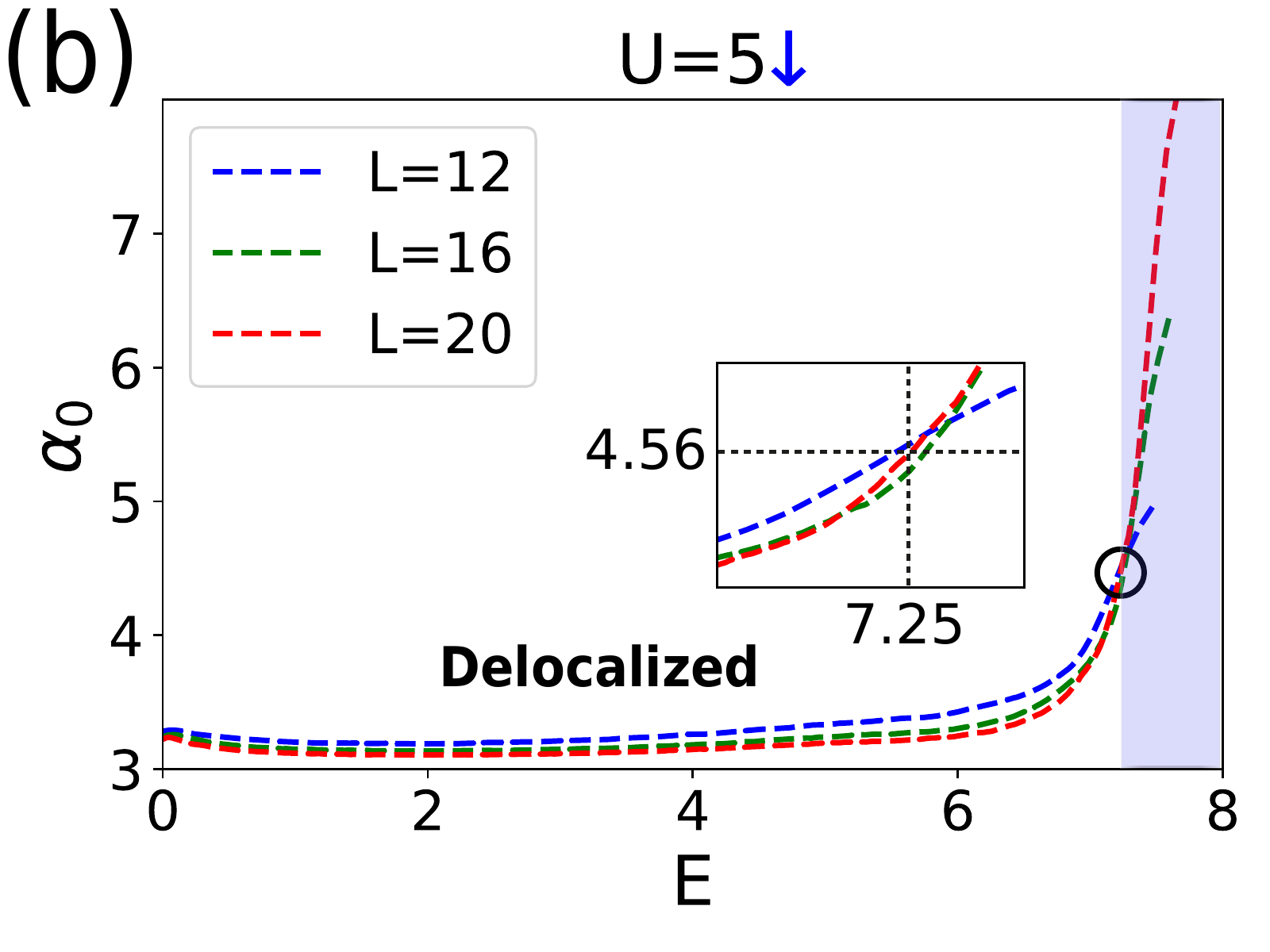}
\includegraphics[scale=0.25]{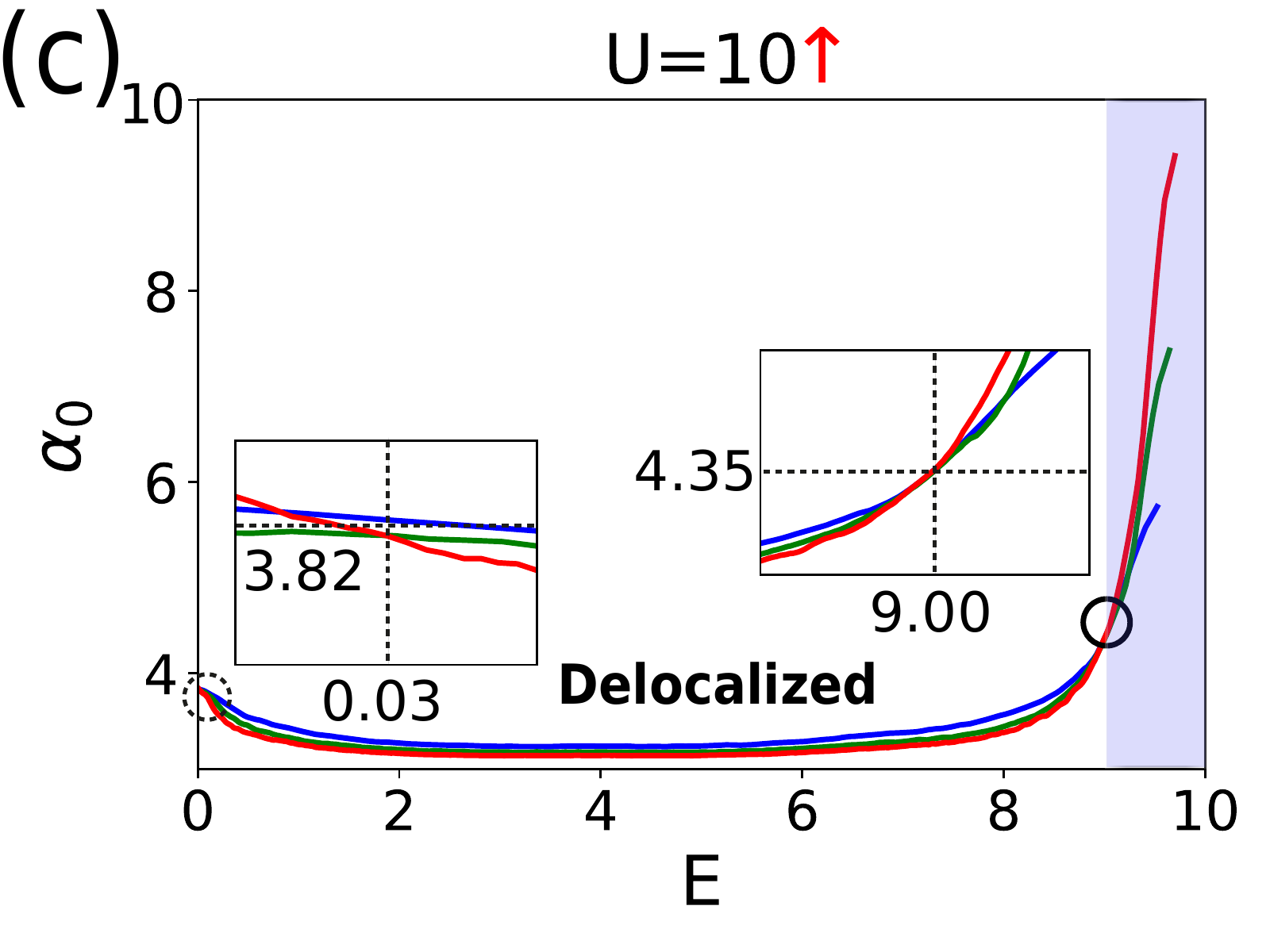}
\includegraphics[scale=0.25]{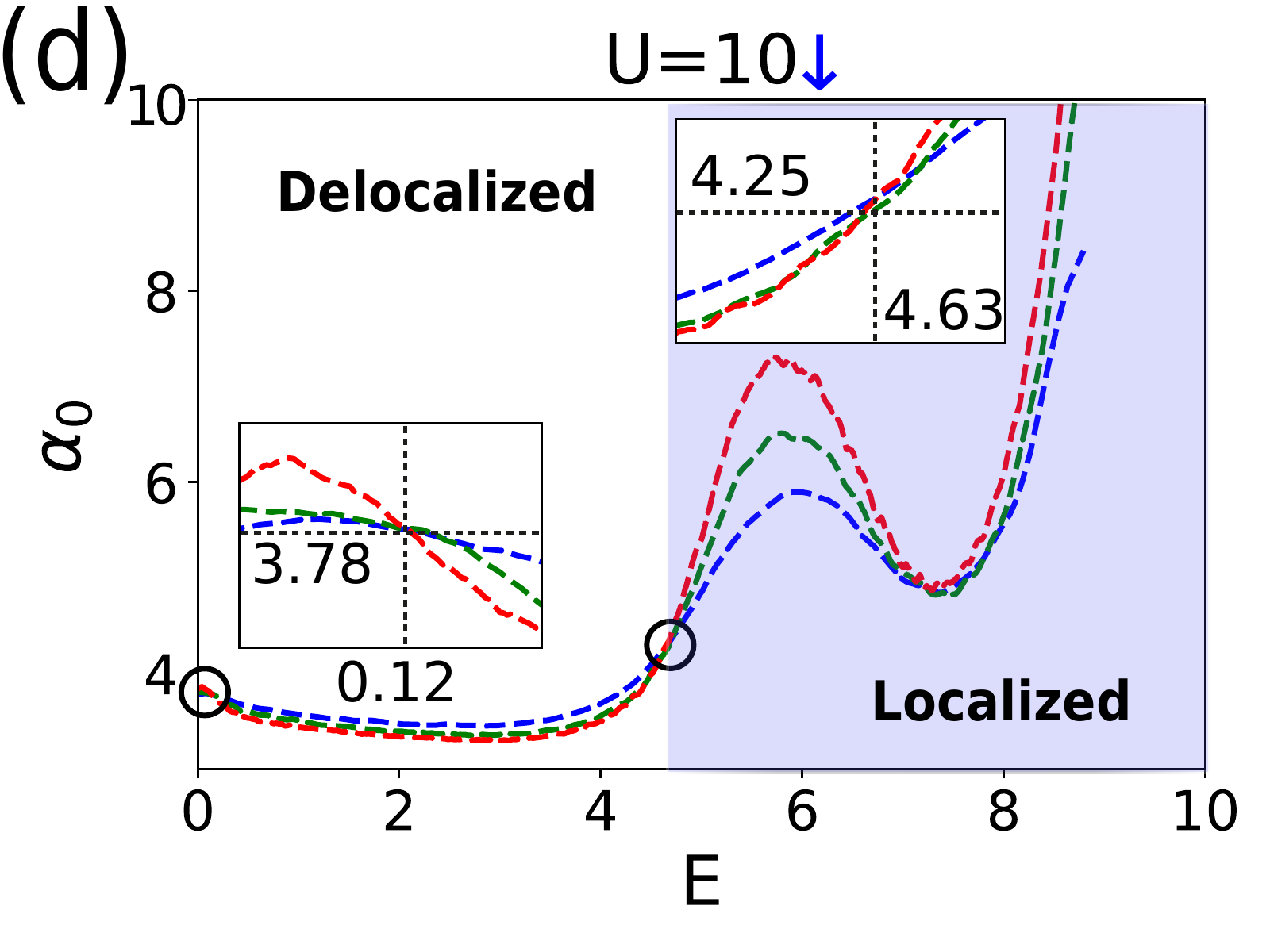}
  \caption{Spin-resolved multifractal scaling exponent $\alpha_0$ for three system sizes $L = 12$, $16$, and $20$ in the SRIS. Here, the left panel shows the multifractal scaling exponent for spin $\uparrow$ electrons, and the right panel represents that for spin $\downarrow$ electrons. Increasing the interaction strength, we find that the low-energy mobility edge arises near the Fermi energy, identified with an Anderson insulating phase. On the other hand, The high-energy mobility edge shows drastic spin resolution in this multifractal scaling exponent, which gives rise to a half metallic phase in an intermediate energy scale, where only spin $\uparrow$ electrons are delocalized. Inset figures clarify the crossing points.}
  \label{fig:Multifractal_Scaling_Exponent_SRIS}
\end{figure}

On the other hand, the multifractal scaling exponent $\alpha_0$ for three system sizes of $L = 12$, $16$, and $20$ in the SRIS shows dramatic spin resolution in Fig. \ref{fig:Multifractal_Scaling_Exponent_SRIS}. The case of $U = 5$ results in a metallic phase at the Fermi energy, where only the high-energy mobility edge is observed to cause an insulating state for the band edge. Here, the dashed line represents the multifractal scaling exponent of the opposite spin flavor. Increasing the interaction strength up to $U = 10$, there appears a crossing point near the Fermi energy, identified with an Anderson insulating phase. One unexpected feature is that the high-energy mobility edge of spin $\downarrow$ electrons in the right panel differs from that of spin $\uparrow$ electrons in the left panel. As a result, only spin $\uparrow$ electrons become metallic in the energy range of $4 < E < 8$ while spin $\downarrow$ electrons are Anderson localized.
%
%

To figure out full information of the eigenfunction statistics, we calculate the multifractal spectrum $f(\alpha)$ for both high energy and low energy mobility edges and both Coulomb and Hubbard interactions. Since the multifractal spectrum is sensitive to the value of the mobility edge, we set the energy window which separates the inversion tendency of the multifractal scaling exponents. Then, the mobility edge is estimated as the
mean value of the energy window. Detailed statistical information is shown in appendix B. Figure \ref{fig:Multifractal_alpha_f_spectrum_SRIS} shows the multifractal spectrum $f(\alpha)$ for all cases just mentioned. Figure \ref{fig:Multifractal_alpha_f_spectrum_SRIS}a and \ref{fig:Multifractal_alpha_f_spectrum_SRIS}b display the multifractal spectrum $f(\alpha)$ of the LRIS ($U = 0.5$) for spin $\uparrow$ and $\downarrow$, while Fig. \ref{fig:Multifractal_alpha_f_spectrum_SRIS}c and \ref{fig:Multifractal_alpha_f_spectrum_SRIS}d show that of the SRIS ($U = 10$) for spin $\uparrow$ and $\downarrow$, respectively. First, it is noticeable that two types of multifractal spectra appear in the LRIS, where blue diamond and red asterisk  points represent the multifractal spectrum $f(\alpha)$ for the high-energy and low-energy mobility edge, respectively. In particular, the multifractal spectrum of the high-energy mobility edge is almost identical to that in the absence of interactions \cite{Two_Mobility_Edges}, denoted by black dots in Fig. \ref{fig:Multifractal_alpha_f_spectrum_SRIS}. On the other hand, the multifractal spectrum of the low-energy mobility edge differs from that of the high-energy mobility edge, which indicates that nature of the metal-insulator transition near the Fermi energy is distinguished from that of the Anderson one without correlations.

Second, the multifractal spectrum for spin $\uparrow$ electrons in the SRIS seems to be quite similar to that of the LRIS in high-energy mobility edge. 
See Figs. \ref{fig:Multifractal_alpha_f_spectrum_SRIS}a, \ref{fig:Multifractal_alpha_f_spectrum_SRIS}b, and \ref{fig:Multifractal_alpha_f_spectrum_SRIS}c. 
On the other hand, the multifractal spectrum for spin $\downarrow$ electrons in the SRIS turns out to be different from that of these cases. As shown in Fig. \ref{fig:Multifractal_alpha_f_spectrum_SRIS}d, the multifractal spectrum of the high-energy mobility edge deviates from that of the Anderson transition without interactions.
This indicates that the metal-insulator transition of spin $\downarrow$ electrons in the SRIS belongs to a novel universality class at least for the high-energy mobility edge, above which the half metallic phase appears to be discussed below.
Although the nature of this metal-insulator transition will be discussed near future, we are strongly suspecting that many-body localization physics \cite{MBL_I,MBL_II} is reflected in the ``quasi-particle" mobility edge. Actually, our recent simulation results based on the multifractal scaling analysis are confirming the existence of a metal-insulator transition at finite temperatures, where the insulating phase persists up to a certain critical temperature. This implies that the corresponding insulating state has to be called a many-body localized phase \cite{MBL_I,MBL_II}. This result is not inconsistent with a novel universality class for a spin-dependent metal-insulator transition in the Hubbard case, as shown in Fig. \ref{fig:Multifractal_alpha_f_spectrum_SRIS}. In this respect we would like to point out that the term of Anderson-type insulator at the Fermi energy should be understood in caution.

\begin{figure}
\includegraphics[scale=0.25]{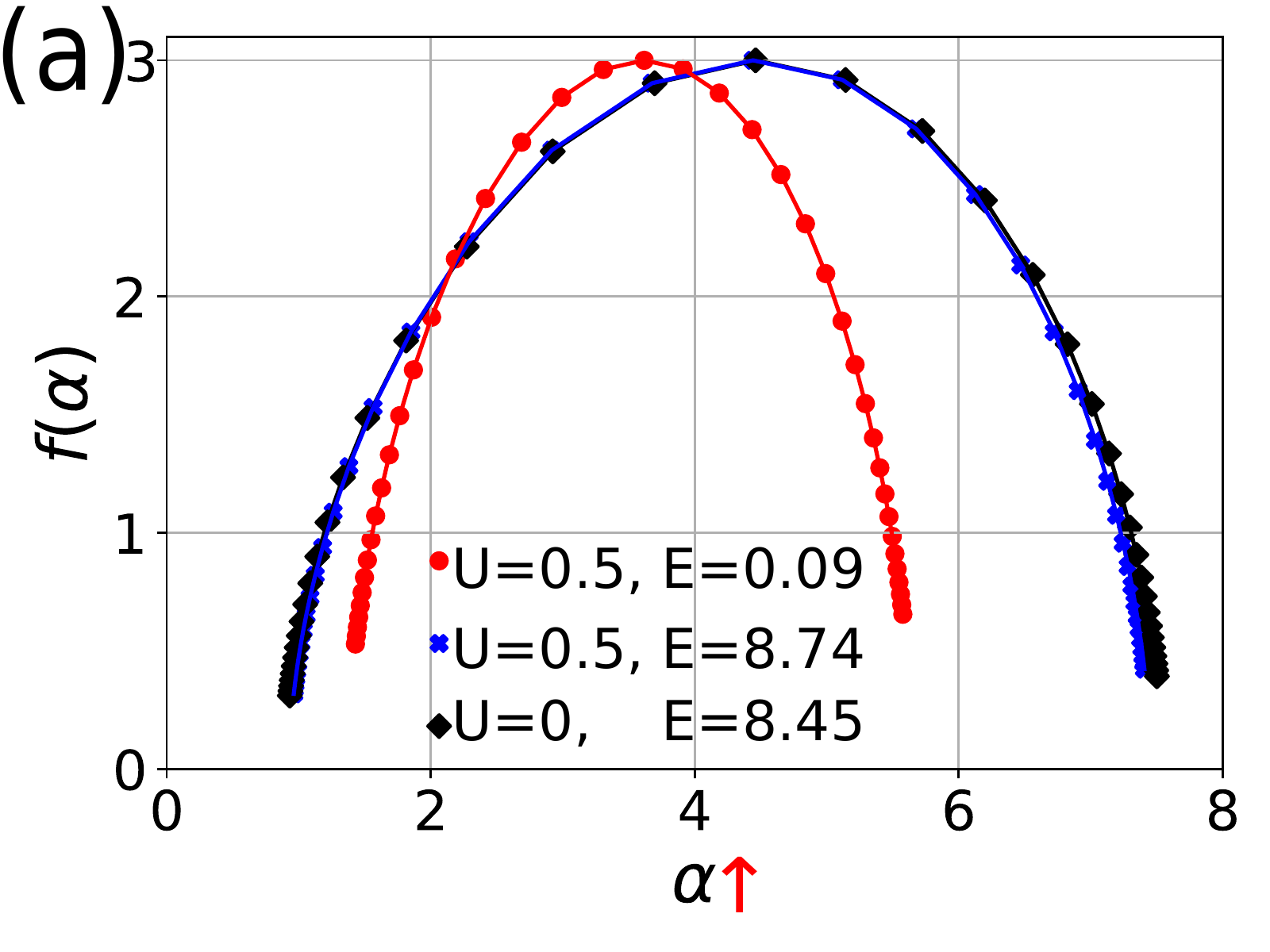}
\includegraphics[scale=0.25]{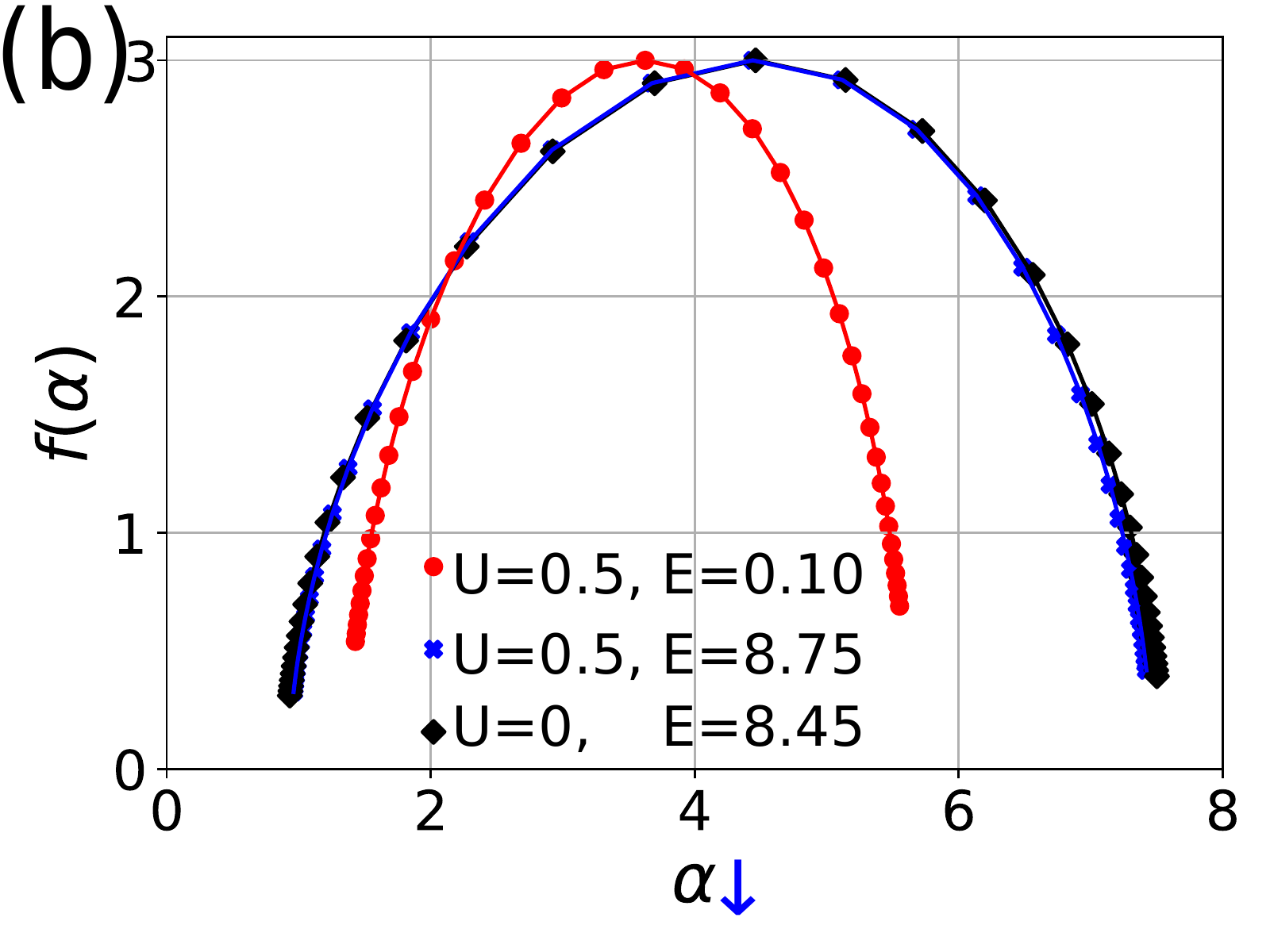}
\includegraphics[scale=0.25]{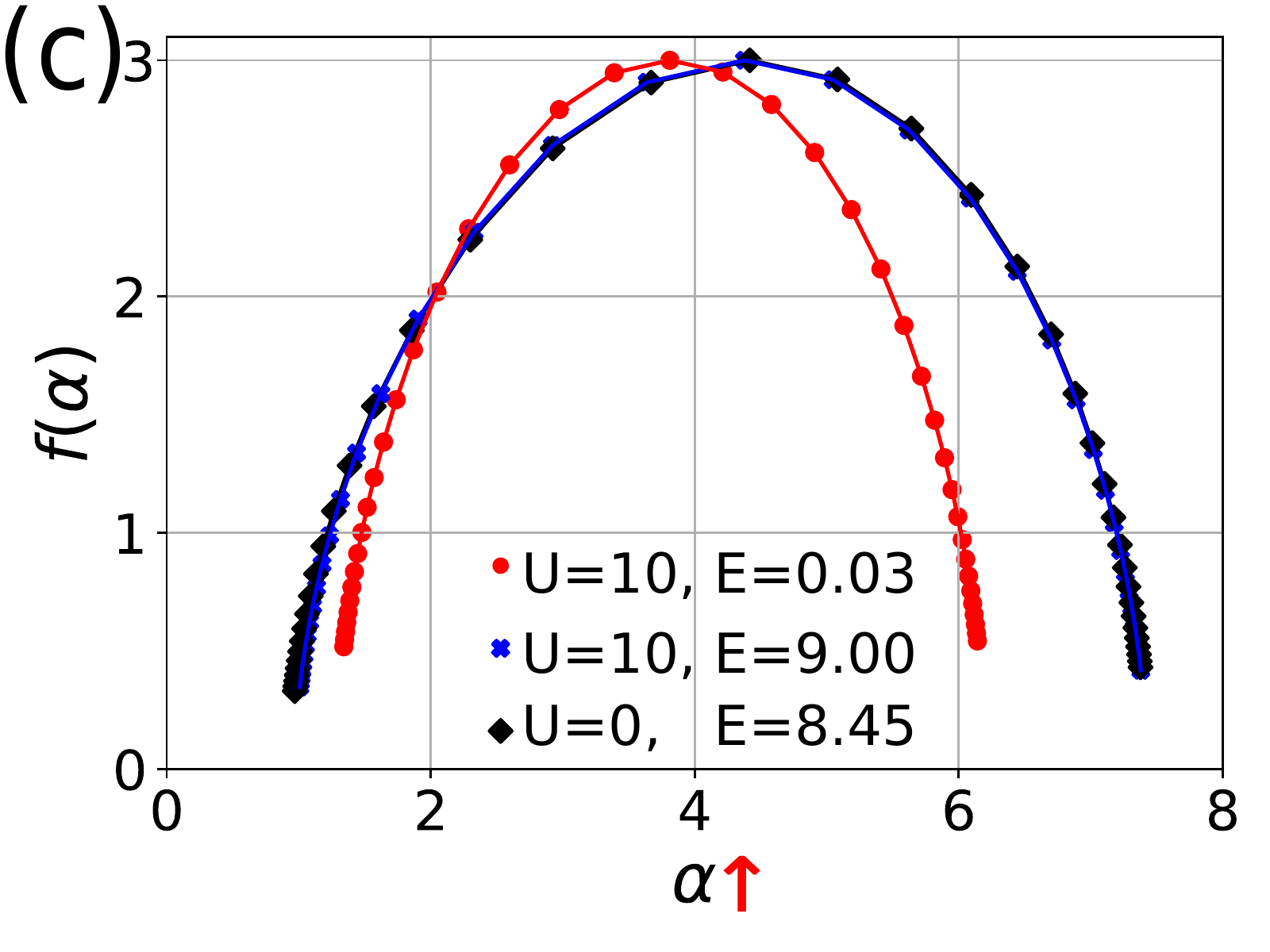}
\includegraphics[scale=0.25]{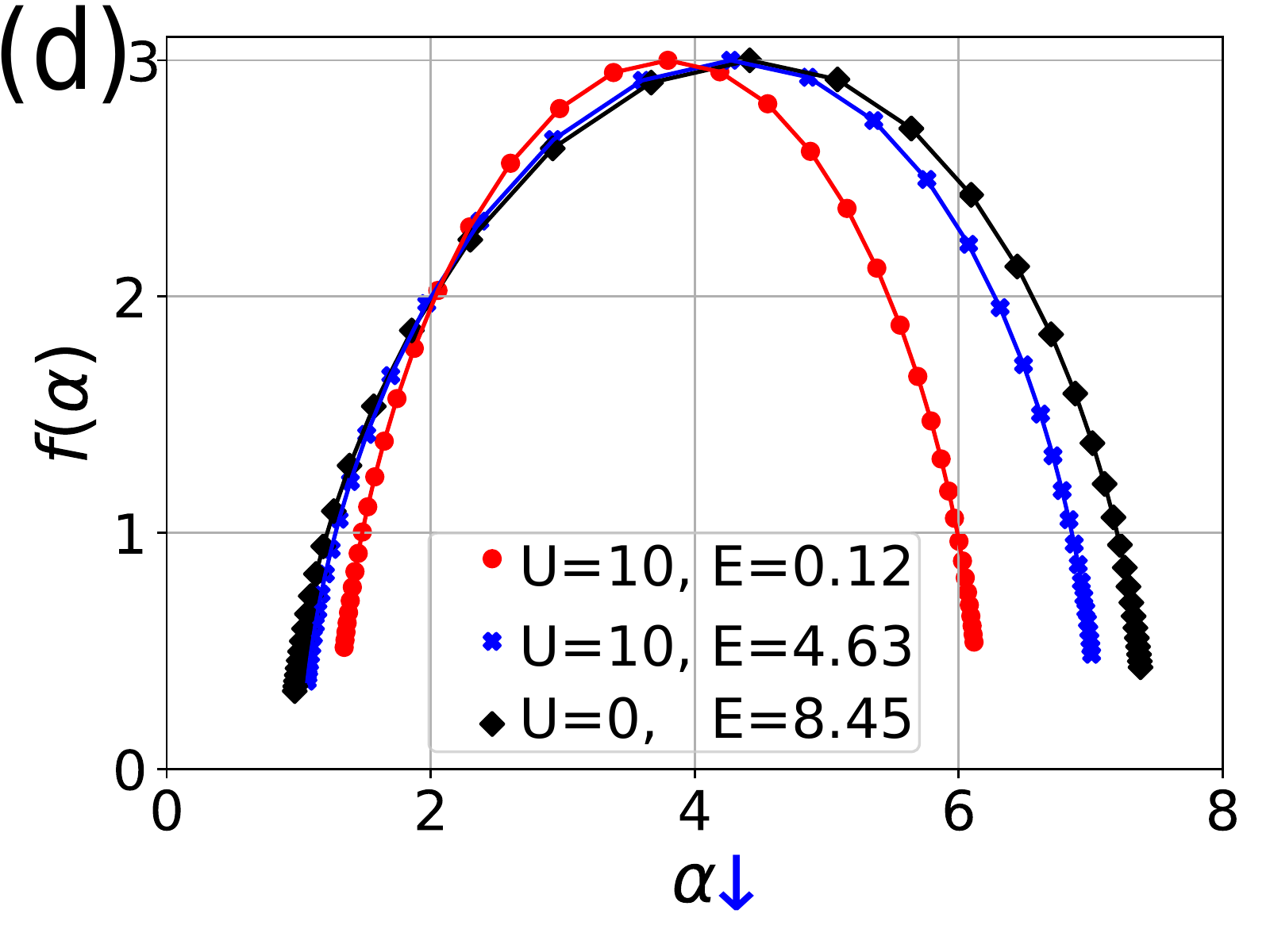}
  \caption{Multifractality singular spectrum for both high-energy and low energy-mobility edges and both Coulomb (a and b) and Hubbard interactions (c and d). See the text for details. Here, blue asterisk (red dots) points denote the multifractal spectrum at the high-energy (low-energy) mobility edge while black diamond represent that of the Anderson transition without correlation effects. It is remarkable to observe that the multifractal spectrum for spin $\uparrow$ electrons in the LRIS is essentially the same as that for spin $\uparrow$ electrons in the SRIS at high-energy mobility edge.
  On the other hand, the multifractal spectrum for spin $\downarrow$ electrons deviates from that of spin $\uparrow$ 
  for the high-energy mobility edge.
  }
  \label{fig:Multifractal_alpha_f_spectrum_SRIS}
\end{figure}

\subsection{Half metals at intermediate energy scales in Anderson insulators}

\begin{figure}
\includegraphics[scale=0.45]{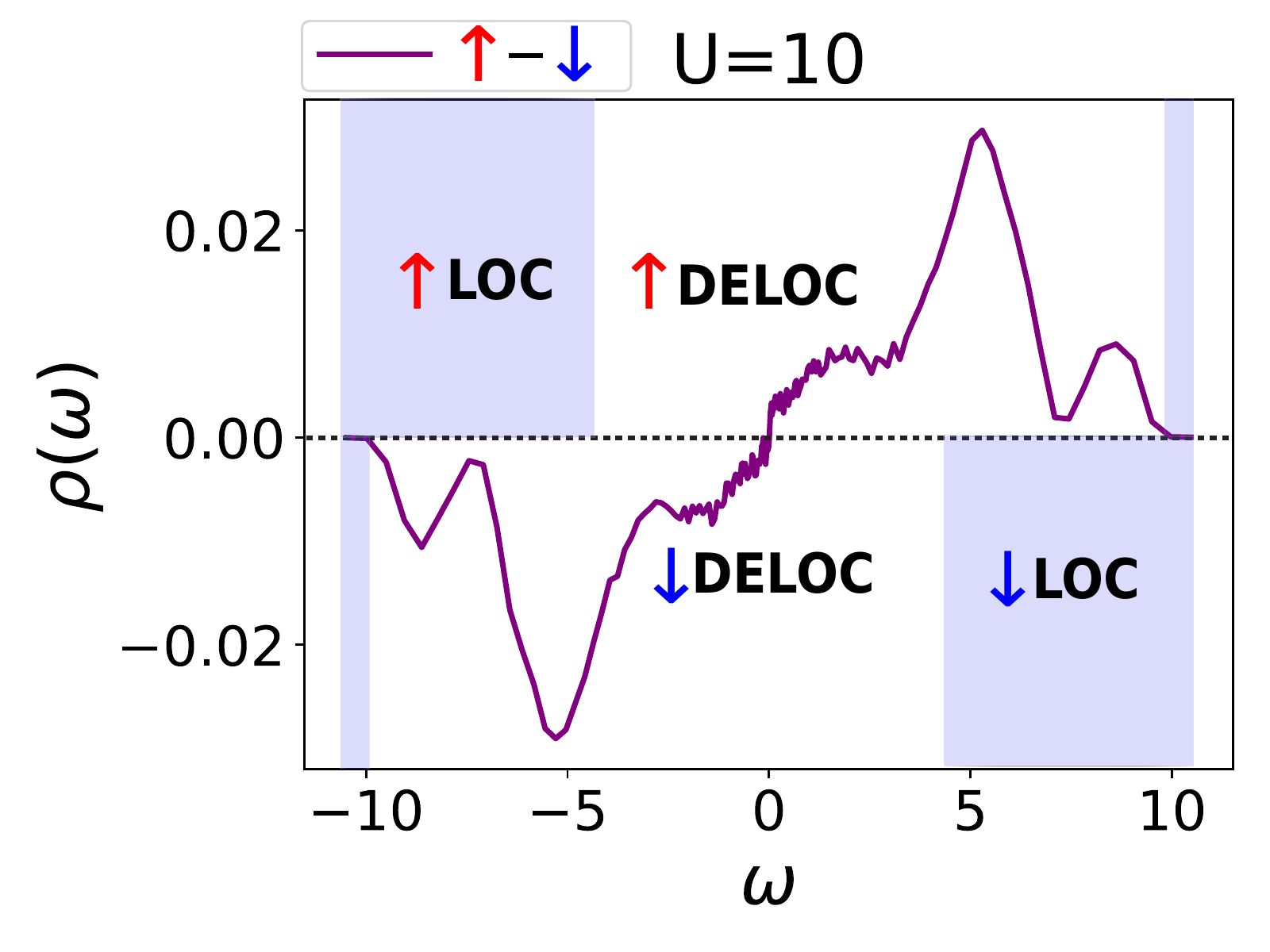}
  \caption{Magnetization density of states $m(\omega) = \rho_{\uparrow}(\omega) - \rho_{\downarrow}(\omega)$ in the SRIS.
  In these interaction-strength ranges, both spin $\uparrow$ and $\downarrow$ electrons near the Fermi energy are Anderson localized. On the other hand, only spin $\uparrow$ ($\downarrow$) electrons are delocalized near the positive (negative) frequency region of the magnetization peak (dip).}
  \label{fig:Magnetization_SRIS}
\end{figure}

Figure \ref{fig:Magnetization_SRIS} shows magnetization density of states $m(\omega) = \rho_{\uparrow}(\omega) - \rho_{\downarrow}(\omega)$ in the SRIS. In these interaction-strength ranges, both spin $\uparrow$ and $\downarrow$ electrons near the Fermi energy are Anderson localized. On the other hand, only spin $\uparrow$ ($\downarrow$) electrons are delocalized near the positive (negative) frequency region of the magnetization peak (dip).
%
%

%
%

%
%

\begin{figure}
\includegraphics[scale=0.5]{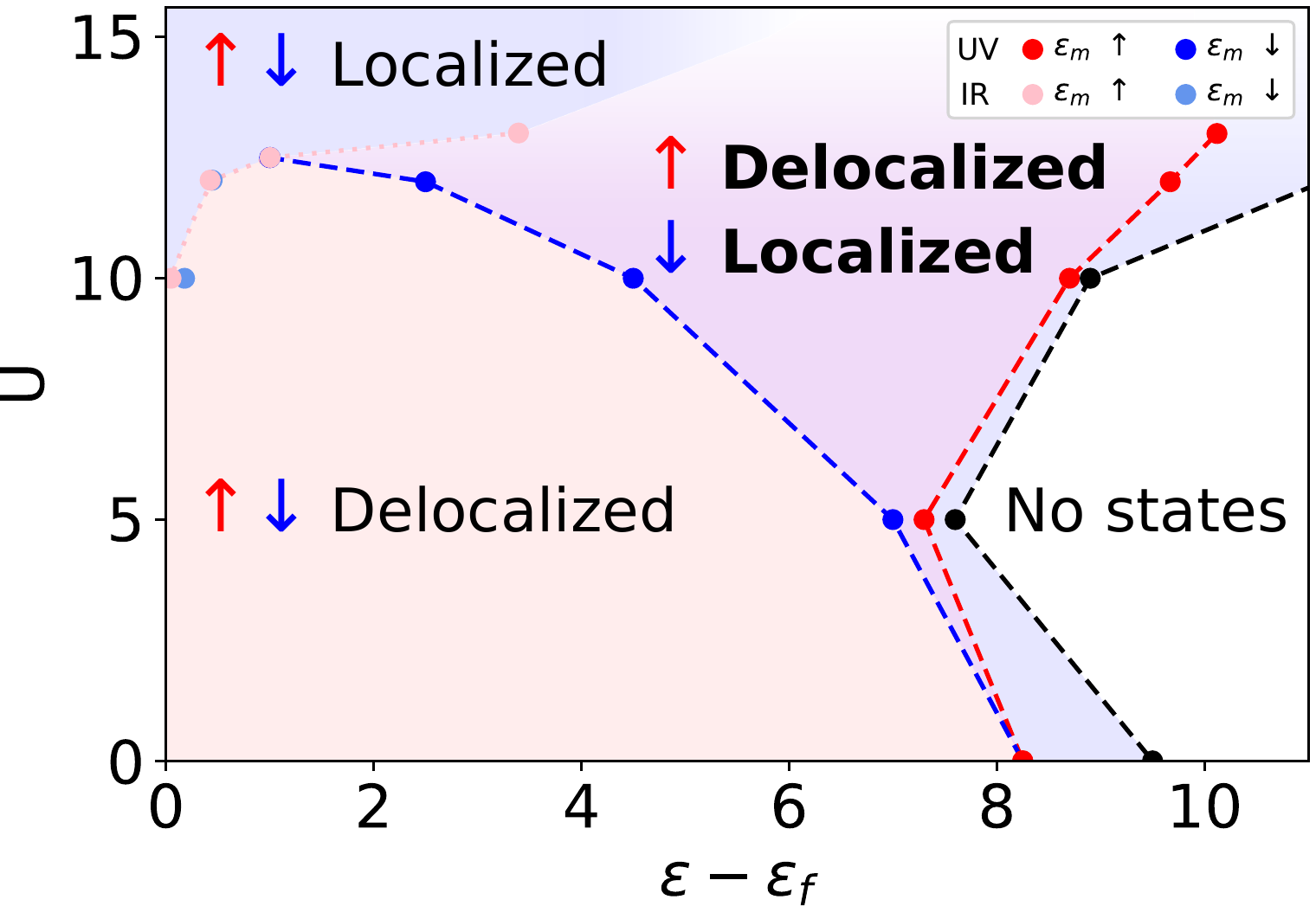}
  \caption{Phase diagram in the plane of energy and interaction strength in the SRIS. Here, we focus on the positive energy side. Red (blue) dots represent the (UV $=$ high-energy and IR $=$ low-energy) mobility edge of spin $\uparrow$ ($\downarrow$) electrons, the above of which corresponds to an Anderson localized phase. There exists a half metallic phase at intermediate energy scales in the Anderson localized state of the Fermi energy.}
  \label{fig:Phase_Diagram_SRIS}
\end{figure}

Based on the multifractal analysis Fig. \ref{fig:Multifractal_Scaling_Exponent_SRIS} with the DOS analysis Fig. \ref{fig:DOS_SRIS} and the magnetization analysis Fig. \ref{fig:Magnetization_SRIS}, we obtain a phase diagram in the plane of energy and Hubbard interaction strength, shown in Fig. \ref{fig:Phase_Diagram_SRIS}. Here, red dots represent the mobility edge for spin $\uparrow$ electrons, and blue ones do that of spin $\downarrow$ electrons. Dashed lines are shown to clarify phase boundaries. When the Hubbard interaction is weak, most energy regions are metallic except for the band edge, given by the high-energy mobility edge. Increasing the interaction strength within the weak-coupling regime in the presence of disorder, where Mott physics is irrelevant (see appendix A), the DOS for spin $\downarrow$ electrons is suppressed at intermediate energy scales while that of spin $\uparrow$ electrons is enhanced at the corresponding energy scales, which results from the role of the Hubbard-type local interactions. As a result, only spin $\uparrow$ electrons are delocalized while spin $\downarrow$ electrons remain to be Anderson localized. Increasing the interaction strength further, the whole band becomes Anderson localized. The emergence of the half metallic state is one of the main results in the present study.

\section{Conclusion}

In this study, we revealed that the role of local magnetic fluctuations in the Anderson metal-insulator transition differs completely between the case of long-ranged Coulomb interactions and that of local Hubbard-type interactions in three dimensions. It turns out that the spin dynamics has nothing to do with the metal-insulator transition in the former. On the other hand, the spin dynamics is essential for the Anderson transition in the latter, where the evolution of the multifractal scaling exponent shows strong spin resolution for both high- and low-energy mobility edges. As a result, we revealed that the multifractal spectrum of spin $\uparrow$ electrons differs from that of spin $\downarrow$ at the high-energy mobility edge, which indicates the existence of spin-dependent universality classes for metal-insulator transitions.
In particular, we found half metals at intermediate energy scales in the Anderson localized phase of the Fermi energy. We believe that the emergence of the half metal phase in Anderson insulators can be verified experimentally by the pump-probe technique \cite{Pump_Probe_Review}.
%
%

Before closing, we would like to discuss why we resort to the multifractal analysis instead of considering the level statistics.
We recall that it is one of the main points the existence of the spin-polarization dependent universality class in the metal-insulator transition,
shown in the multifractal spectrum of Fig. \ref{fig:Multifractal_alpha_f_spectrum_SRIS}.
We point out that a spin-resolved metal-insulator transition would
appear in the level statistics analysis \cite{Level_Statistics}. In addition, showing that the critical exponent of the localization length for spin $\uparrow$ electrons differs from that for spin $\downarrow$ ones, one may claim that this spin-resolved metal-insulator transition separates into two universality classes.
On the other hand, we would like to point out that the difference of the multifractal spectrum between spin $\uparrow$ and $\downarrow$ electrons is not so serious, as shown in the multifractal spectrum. Although we did not evaluate the localization length exponent in this study, the information from the multifractal spectrum seems to imply that both critical exponents might be close, giving rise to some obstructions in claiming the existence of two universality classes. We would like to leave the study based on the level statistics and the localization length as an interesting future direction.

\acknowledgments

This study was supported by the Ministry of Education, Science, and Technology (No. 2018R1A5A6075964 and NRF-2021R1A2C1006453) of the National Research Foundation of Korea (NRF).

\appendix

\section{Justification of the Hartree-Fock-Anderson analysis}

First, let us clarify essential ingredients in our Hartree-Fock-Anderson analysis, comparing this method with the Finkelstein's renormalization group (RG) analysis \cite{Finkelstein_RG}. The Finkelstein's RG analysis takes into account both effects of (weak) localization and (weak) interaction self-consistently based on an effective field theory, referred to as nonlinear $\sigma$ model. This effective field theory deals with self-consistent renormalization effects of diffusions and Cooperons for weak localization and spin-singlet and spin-triplet collective excitations for weak interaction. The existence of a metal-insulator transition in two spatial dimensions has been argued in the one-loop level and confirmed in the two-loop level \cite{Finkelstein_RG}. On the other hand, the Hartree-Fock-Anderson analysis takes into account these interaction effects of spin-singlet and spin-triplet in the one-loop level, but deals with localization effects essentially exactly via brute force disorder averaging. In this respect the Hartree-Fock-Anderson analysis describes not only physics of a metallic state but also that of an insulating phase beyond the Finkelstein's RG analysis that cannot cover the insulating regime explicitly. Furthermore, our Hartree-Fock-Anderson framework can be extended to introduce interaction effects of the two-loop order in a straightforward manner, where $1/N$ quantum corrections ($N$ is the spin degeneracy) are given by ``Goldstone" excitations in real space. More precisely, such quantum corrections are $\sim \sum_{\bm{r}_{i}} \sum_{i\Omega} \ln \Pi(\bm{r}_{i}, \bm{r}_{i}, i\Omega)$ in the Hartree-Fock-Anderson free-energy functional of order parameters, where the fully renormalized polarization function is given by $\Pi(\bm{r}_{i}, \bm{r}_{j}, i\Omega) \sim N \sum_{\bm{r}_{k}} \sum_{i\omega} G(\bm{r}_{i}, \bm{r}_{k}, i\omega+i\Omega) G(\bm{r}_{k}, \bm{r}_{j}, i\omega)$ with the fully renormalized Hartree-Fock-Anderson single-particle Green's function $G(\bm{r}_{k}, \bm{r}_{j}, i\omega)$.

Armed with this argument, second, we compare our Hartree-Fock-Anderson analysis with Phys. Rev. B \textbf{78}, 235116 (2008) \cite{Henseler}, and confirm that the half metallic state appears robustly in the weak coupling regime. Ref. \cite{Henseler} proposed an interesting hybridization framework between the coherent potential approximation (CPA) for the description of a Mott insulating phase and the Woelfle-Vollhardt self-consistent theory for Anderson localization \cite{Self_consistent_Theory_AL}. This study describes the appearance of Mott gap in the density of states as increasing the Hubbard interaction based on the CPA framework with renormalized chemical potentials, which reproduces a Mott insulating phase in the atomic limit. Resorting to this single-particle Green's function, this previous investigation constructs a self-consistent theory for Anderson localization, which extends the Woelfle-Vollhardt self-consistent theory \cite{Self_consistent_Theory_AL} without correlation effects up to an effective theoretical framework with Mott gap.

\begin{figure}
\includegraphics[scale=0.25]{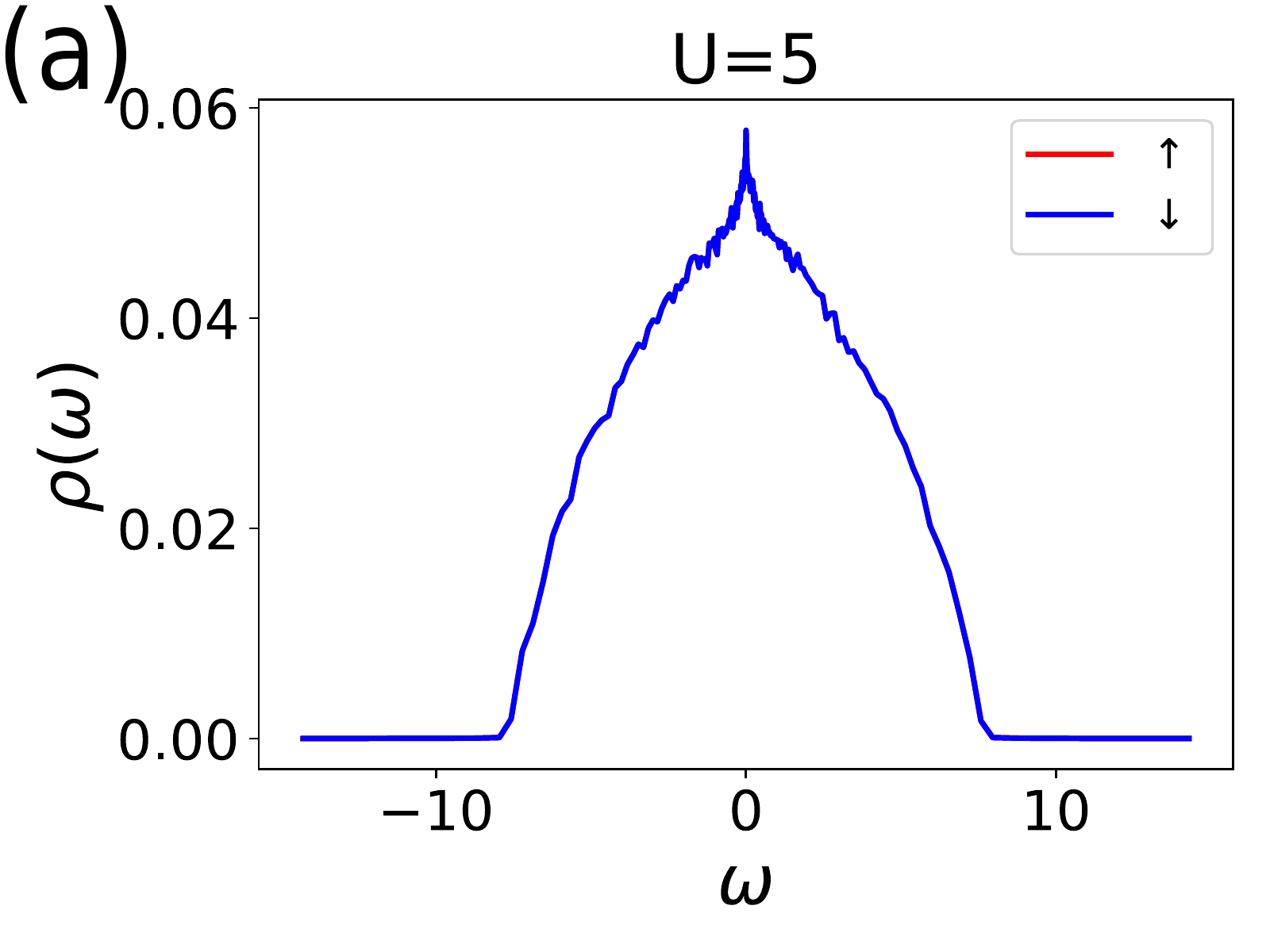}
\includegraphics[scale=0.25]{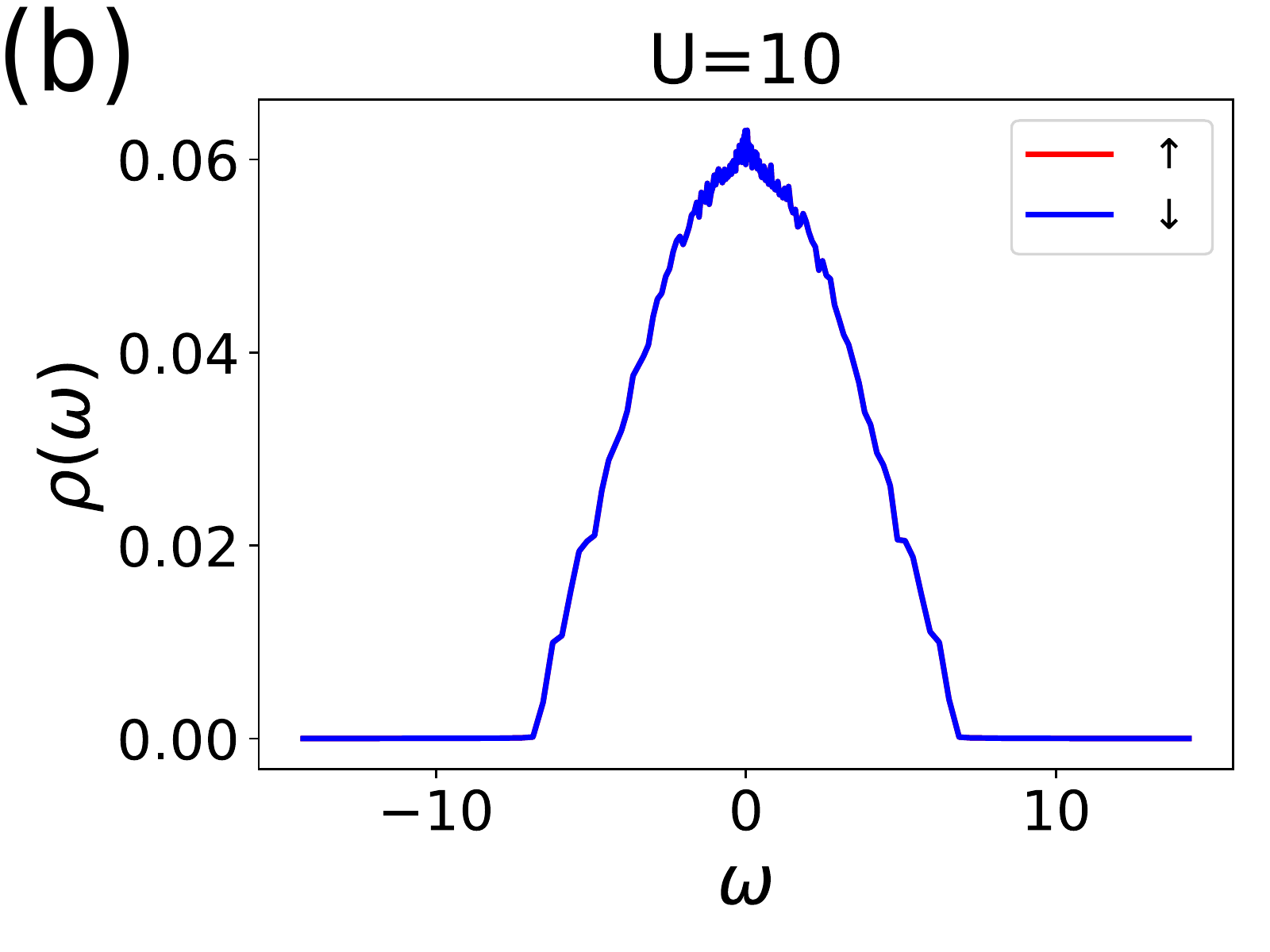}
  \caption{Disorder averaged density of states (DOS) $\rho(\omega)$ for various interaction parameters in the short-ranged interacting system (SRIS). These results based on the Hartree-Fock-Anderson analysis are consistent with those [Fig. 1 of Ref. \cite{Henseler}] of the CPA framework up to the weakly interacting regime $U \leq 10$.}
  \label{fig:Dos_spinless}
\end{figure}

To compare these two methods with each other, we first consider a paramagnetic phase ($\langle n_{i \uparrow} \rangle = \langle n_{i \downarrow} \rangle = \frac{\langle n_{i \uparrow} \rangle + \langle n_{i \downarrow} \rangle}{2}$) in the Hartree-Fock-Anderson analysis. Although the Hartree-Fock-Anderson analysis fails to generate the Mott gap, the gapless feature of the density of states in the Hartree-Fock-Anderson analysis turns out to be qualitatively similar to that of the CPA framework up to $U/t = 8.5$ [Fig. 1(c) of Ref. \cite{Henseler}]. See Fig. \ref{fig:Dos_spinless}. On the other hand, the dip feature at the Fermi energy clearly shown in Fig. 1(c) of Ref. \cite{Henseler} starts to appear at larger $U/t$, at least $U/t > 10$ in the Hartree-Fock-Anderson analysis. Interestingly, this paramagnetic-solution demonstration in the Hartree-Fock-Anderson analysis indicates that the dip feature at the Fermi energy, which occurs at smaller values of $U/t$, results from spin polarization with the diffusive nature of electrons at the Fermi energy. More importantly, as clearly shown in the phase diagram of the manuscript, the half metallic state appears in the weak coupling regime. More precisely, focusing on the energy cut $\varepsilon - \varepsilon_{F} = 7$, for example, the half metallic phase starts from $U/t \approx 4$ and persists up to more than $U/t = 10$, at least in the parameter regime between which the Hartree-Fock-Anderson analysis is physically justified. See the phase diagram Fig. 8 of the main text.

In summary, (i) we argued that the present Hartree-Fock-Anderson analysis is justified at least in the weak coupling regime as much as the Finkelstein's RG analysis, and (ii) claimed that the half metallic state at an intermediate energy scale indeed occurs in the weak coupling regime, verified in comparison with Ref. \cite{Henseler}.
\\

\section{Statistical information for the multifractal spectrum of Fig. \ref{fig:Multifractal_alpha_f_spectrum_SRIS}}

To obtain the multifractal spectrum, we estimated the mobility edge as an average value of the energy window that separates the inversion tendency of the multifractal scaling exponents. The number of states in this energy window ranges from $1,000$ to $5,000$ for the SRIS and from $200$ to $1,000$ for the LRIS. The range of the mobility edge, the mean value, and the standard deviation are shown in table \ref{table1} and \ref{table2}. This averaging procedure results in the maximum standard error $0.023$ for $\alpha_q$ and $0.013$ for $f(\alpha_q)$, respectively, where $-5<q<5$.

\begin{table}[h!]
\centering
\begin{tabular}{|c|c|c|c|}
\cline{1-4}
  &$\delta E_m$ & $\overline{E_m}$ & $\sigma_{E_m}$ \\\cline{1-4}

     $\uparrow IR$ & $0.04<E_m<0.14$& $0.09$ & 0.05
     \\\cline{1-4}
$\uparrow UV$ & $8.65<E_m<8.85$& $8.74$ & 0.07
     \\\cline{1-4}
     $\downarrow IR$ & $0.05<E_m<0.15$& $0.11$ & 0.06
     \\\cline{1-4}
     $\downarrow UV$ & $8.65<E_m<8.85$& $8.75$ & 0.07
     \\\cline{1-4}
     \end{tabular}
      \caption{Statistics of the mobility edges for U=0.5 in the LRIS.}
     \label{table1}
\end{table}

\begin{table}[h!]
\centering
\begin{tabular}{|c|c|c|c|}
\cline{1-4}
  &$\delta E_m$ & $\overline{E_m}$ & $\sigma_{E_m}$ \\\cline{1-4}
  $\uparrow IR$ & $0<E_m<0.05$& $0.03$ & 0.10
     \\\cline{1-4}
$\uparrow UV$ & $8.9<E_m<9.1$& $9.00$ & 0.09
     \\\cline{1-4}
     $\downarrow IR$ & $0.10<E_m<0.15$& $0.12$ & 0.08
     \\\cline{1-4}
     $\downarrow UV$ & $4.54<E_m<4.74$& $4.63$ & 0.14
     \\\cline{1-4}

     \end{tabular}
      \caption{Statistics of the mobility edges for U=10 in the SRIS.}
     \label{table2}
\end{table}
%
%


\begin{thebibliography}{99}
\bibitem{Comment_Ewald_Sum} In order to clarify the role of long-range interactions in the Hartree-Fock approximation, we implement the Ewald summation technique, where the Hartree term is split into two parts: a real-space portion based on a short-range interaction potential whose pairwise sum converges quickly and a long-range portion based on a slowly-varying interaction potential whose pairwise sum converges relatively quickly in a reciprocal space \cite{Ewald_Sum_Technique1,Ewald_Sum_Technique2,Ewald_Sum_Technique3}. The optimal implementation of the Ewald technique allows us to resolve the long-standing issue of the ill-convergence of the long-range potential and the multiplicity of Hartree-Fock solutions near the Anderson-Mott transition, which has been also reported in a recent Hartree-Fock numerical study \cite{Multifractality_Interaction_Numerics}.
\end{thebibliography}
\end{document}